\documentclass[conference]{IEEEtran}
\IEEEoverridecommandlockouts
\usepackage{cite}
\usepackage{algorithmic}
\usepackage{graphicx}
\usepackage{xcolor}
\usepackage[linesnumbered,ruled,noend]{algorithm2e} 
\usepackage{multirow}
\usepackage{url}
\usepackage{bm}
\usepackage{amsmath,amssymb,amsfonts}

\usepackage{textcomp}
\usepackage{amsthm}
\usepackage{subfigure}
\usepackage{paralist}
\usepackage{stfloats}
\usepackage{bbm}
\usepackage{siunitx}
\usepackage{hhline}
\usepackage{colortbl}
\usepackage{wrapfig}
\usepackage{makecell}
\usepackage{booktabs}

\def\BibTeX{{\rm B\kern-.05em{\sc i\kern-.025em b}\kern-.08em
    T\kern-.1667em\lower.7ex\hbox{E}\kern-.125emX}}
\begin{document}

\title{TimeKit: A Time-series Forecasting-based Upgrade Kit for Collaborative Filtering}

\author{\IEEEauthorblockN{Seoyoung Hong$^*$, Minju Jo$^*$, Seungji Kook$^*$, Jaeeun Jung$^\dagger$, Hyowon Wi$^*$, Noseong Park$^*$, Sung-Bae Cho$^*$}
\textit{Yonsei University$^*$, Seoul, South Korea}\\
\textit{Kakao Corporation$^\dagger$, Seoul, South Korea}\\
\{seoyoungh, alflsowl12, 2021321393, wihyowon, noseong, sbcho\}@yonsei.ac.kr$^*$, \{jeni.th\}@kakaocorp.com$^\dagger$}


\IEEEoverridecommandlockouts
\IEEEpubid{\makebox[\columnwidth]{978-1-6654-8045-1/22/\$31.00~\copyright2022 IEEE \hfill}
\hspace{\columnsep}\makebox[\columnwidth]{ }}

\maketitle

\IEEEpubidadjcol

\begin{abstract}
Recommender systems are a long-standing research problem in data mining and machine learning. They are incremental in nature, as new user-item interaction logs arrive. In real-world applications, we need to periodically train a collaborative filtering algorithm to extract user/item embedding vectors and therefore, a time-series of embedding vectors can be naturally defined. We present a time-series forecasting-based upgrade kit (TimeKit), which works in the following way: it i) first decides a base collaborative filtering algorithm, ii) extracts user/item embedding vectors with the base algorithm from user-item interaction logs incrementally, e.g., every month, iii) trains our time-series forecasting model with the extracted time-series of embedding vectors, and then iv) forecasts the future embedding vectors and recommend with their dot-product scores owing to a recent breakthrough in processing complicated time-series data, i.e., neural controlled differential equations (NCDEs). Our experiments with four real-world benchmark datasets show that the proposed time-series forecasting-based upgrade kit can significantly enhance existing popular collaborative filtering algorithms.
\end{abstract}


\section{Introduction}
Recommender systems, personalized information filtering (IF) technologies, can be applied to many services, ranging from E-commerce, advertising, and social media to many other online and offline service platforms~\cite{covington2016deep, Ying18pinsage,Yang2016}. One of the most popular recommender systems, collaborative filtering (CF), provides personalized preferred items by learning user and item embeddings from user-item interactions~\cite{8506344, Ebesu2018, He17NeuMF, Hu08WRMF, 5197422, Rendle09BPR, Wang2014}.

\begin{figure}
    \centering
    \subfigure[Existing concept where $\mathcal{A}$ embeds users/items onto a vector space]{\includegraphics[width=0.75\columnwidth]{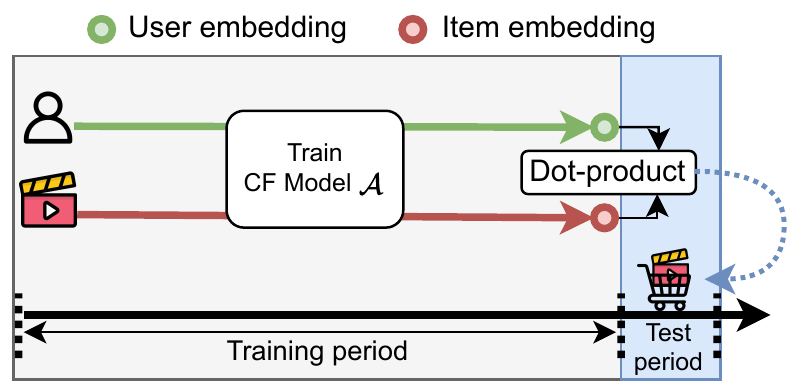}}    \subfigure[Forecasting the future embedding vectors via time-series forecasting technology, where the training period can span multiple years]{\includegraphics[width=0.75\columnwidth]{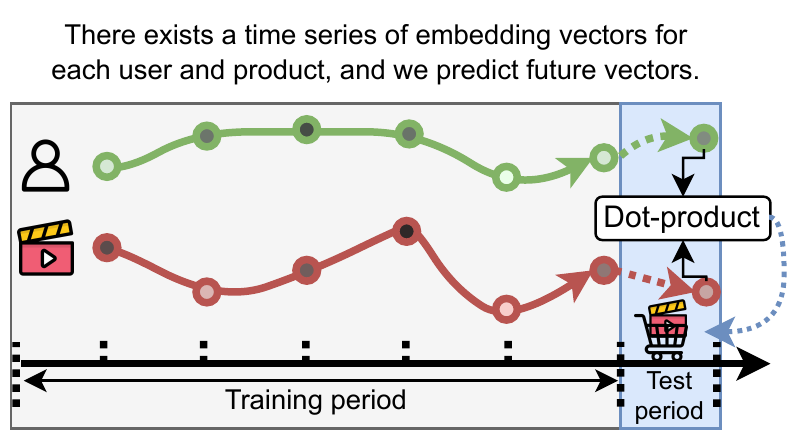}}
    \caption{The comparison between our proposed and existing methods: (a) Existing recommendation technologies train embedding vectors with collected data. (b) We design an advanced time-series forecasting model to forecast the future embedding vectors of users/items.}
    \label{fig:teaser}
\end{figure}

However, most of collaborative filtering methods do not consider the practical point that real-world applications require re-training them periodically --- existing collaborative filtering methods consider one-time training only as in Fig.~\ref{fig:teaser} (a). Let $\mathcal{A}$ be a base collaborative filtering algorithm which produces the embedding vectors, denoted $\{\bm e_{u}^{Last}\}_{u=1}^{|U|}$, $\{\bm e_{v}^{Last}\}_{v=1}^{|V|}$ where $U$ (resp. $V$) is a set of users (resp. items), at its last layer. In general, $\mathcal{A}$ calculates the dot-product values of the embedding vectors for recommendation. In real-world environments, we need to train $\mathcal{A}$ periodically as new user-item interactions are incremental. As a result, one may naturally construct the time-series of user/item embedding vectors. Owing to the recent advancements in time-series forecasting, We propose forecasting future user/item embedding vectors which describe latent behavioral patterns. (cf. Fig.~\ref{fig:teaser} (b)). The key in forecasting the future embedding vectors is understanding the underlying dynamics which causes the behavioral drift --- in other words, user behavioral patterns change over time. To address this, we resort to recent time-series forecasting technologies. We will show that our proposed method can drastically enhance existing popular collaborative filtering algorithms, both older MF and newer GCN-based models. Therefore, we call it as a time-series forecasting-based upgrade kit of collaborative filtering (TimeKit).

Let $\mathcal{E}_i=\{\bm e_{u}^{Last}\}_{u=1}^{|U|} \cup \{\bm e_{v}^{Last}\}_{v=1}^{|V|}$ be a set of all embedding vectors by the base algorithm $\mathcal{A}$ trained with the data collected up to a time-point $t_i$.
In our problem setting, our forecasting model, given a time-series of embedding vectors $\{(\mathcal{E}_i, t_i)\}_{i=1}^M$, forecasts the future embedding vectors $\mathcal{E}_{M+1}$ that hopefully work well for the future time period $(t_M, t_{M+1}]$.

\begin{figure}
    \centering
    \subfigure[Recall@20]{\includegraphics[width=0.49\columnwidth]{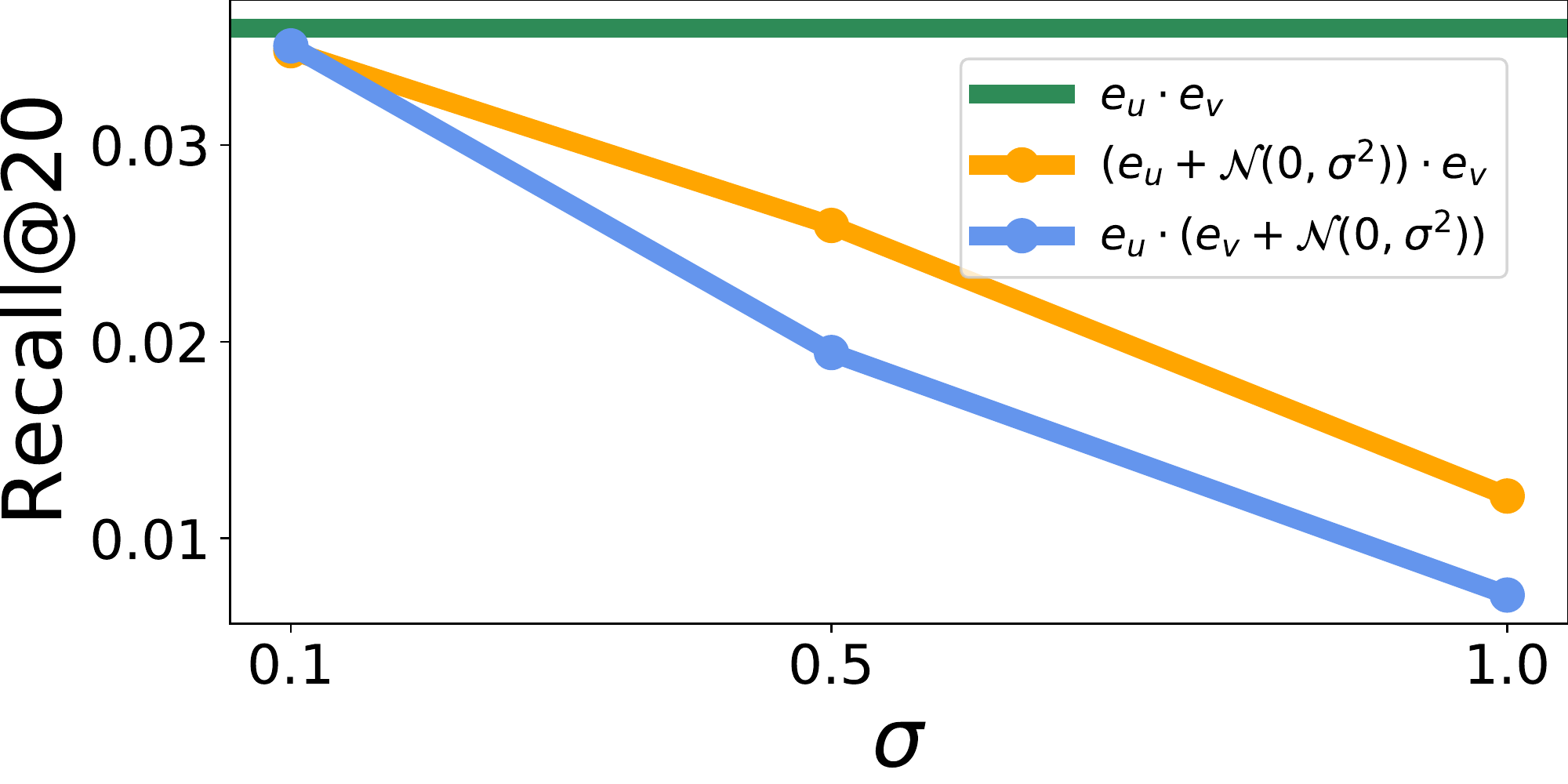}}
    \subfigure[NDCG@20]{\includegraphics[width=0.49\columnwidth]{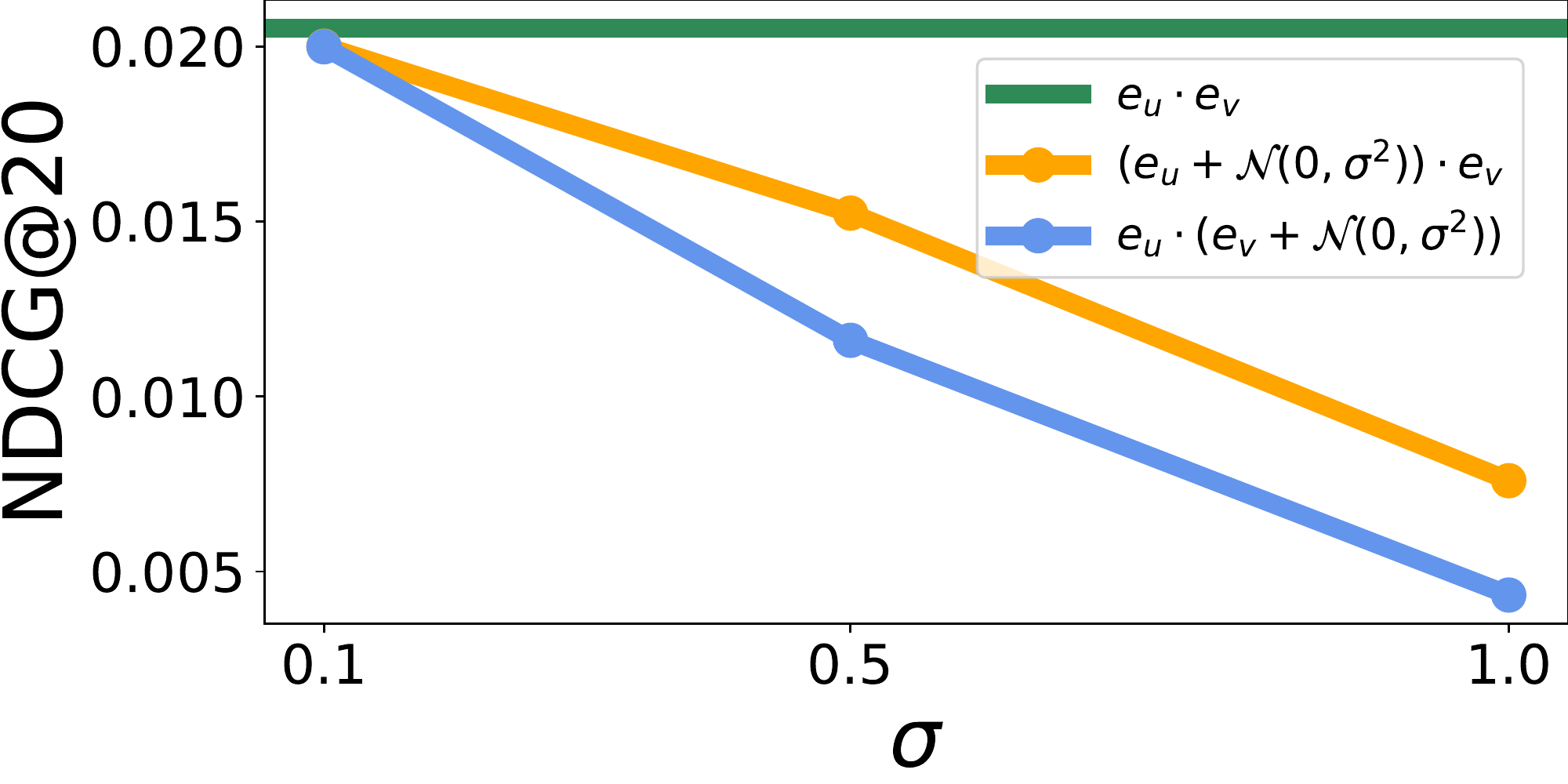}}\hfill
    
    \caption{We compare three cases in Goodreads: i) the dot-product between original user and item embedding vectors, denoted $\bm{e}_u \cdot \bm{e}_v$, calculated by LightGCN, one of the state-of-the-art collaborative filtering algorithms, ii) the dot-product between original user and perturbed item embedding vectors, denoted $\bm{e}_u\cdot (\bm{e}_v+\mathcal{N}(0,\sigma^2))$, and iii) the dot-product between perturbed user and original item embedding vectors, denoted $(\bm{e}_u+\mathcal{N}(0,\sigma^2)) \cdot \bm{e}_v$. As noted, item embedding vectors are more sensitive to $\sigma$.}
    \label{fig:error}
\end{figure}


Our proposed approach will predict the future interactions better than existing methods when the future embedding vectors are correctly predicted. However, this forecasting task is challenging. We found that time-series of user embedding vectors can be better processed by gated recurrent units (GRUs) and those of item embedding vectors can be better processed by neural controlled differential equations (NCDEs~\cite{NEURIPS2020_4a5876b4}) --- after analyses, we found that in general, item embedding vectors are more sensitive to small errors than user embedding vectors and are more challenging to forecast (cf. Fig.~\ref{fig:error}). NCDEs can be used to forecast item embedding vectors, which requires more accuracy. Therefore, a specific combination of GRU and NCDE shows the best performance in our experiments in many cases. NCDEs are considered as a \emph{continuous} analogue to RNNs and written as follows:
\begin{align}\label{eq:ncde}
\bm{z}(T) &= \bm{z}(0) + \int_{0}^{T} f(\bm{z}(t);\bm{\theta}_f) dX(t)\\&= \bm{z}(0) + \int_{0}^{T} f(\bm{z}(t);\bm{\theta}_f) \frac{dX(t)}{dt} dt,\label{eq:ncde2}
\end{align}where $\bm{z}(t)$ is a vector at time $t$, and $X$ is a continuous path taking values in a Banach space. The theory of the controlled differential equation (CDE) had been developed to extend the stochastic differential equation and the It\^{o} calculus far beyond the semimartingale setting of $X$~\cite{protter1985approximations,cont2013functional}. For instance, a prevalent example of the path $X$ is a Wiener process, in which case Eq.~\eqref{eq:ncde} reduces to a stochastic differential equation. In CDEs, however, the path $X$ does not need to be such semimartingale or martingale processes. These CDEs are actively utilized in financial markets to predict future financial environments influenced by uncertainties. NCDEs are technologies to parameterize such CDEs and learn from data. In addition, Eq.~\eqref{eq:ncde2} continuously reads the values $\frac{dX(t)}{dt}$ and integrates them over time. In this regard, NCDEs are equivalent to continuous RNNs and show the state-of-the-art performance in many time-series tasks and datasets. The reasons of our adopting NCDEs are as follows:
\begin{compactenum}
\item There are several existing results showing that differential equation-based models extrapolate (forecast) better than RNNs for the data in the field of social science, physical science, finance, and so forth~\cite{10.2307/1831029,erichson2019physics,raissi2019physics,ZHANG2019108850,kim2020dpm}. Recall that the concept of CDEs were initially created to describe the uncertainties in financial markets. We conjecture that this is likely to be the case in our work as well (because our work also deals with human behavioral patterns).
\item How to define the CDE function $f$ in Eq.~\eqref{eq:ncde} is critical in terms of forecasting accuracy and easiness of training --- note that $f$ is a neural network parameterized by $\bm{\theta}_f$. We use only $\emph{Lipschitz}$ operators for designing $f$ to make the problem of finding the optimal parameter $\bm{\theta}^*_f$ \emph{well-posed} (see the discussion in Section~\ref{sec:tr}).
\end{compactenum}

We do experiments with 4 real-world datasets. Since solving a novel task, we define some baselines and compare our method with them. Our upgrade kit enhances the base algorithm by up to 68.46\% for Recall@20 and up to 68.75\% for NDCG@20. Our contributions are summarized as follows:
\begin{compactenum} 
    \item We reduce the collaborative filtering (CF) problem to a time-series forecasting of embedding vectors and design an upgrade kit of CF algorithms, called \emph{TimeKit}.
    \item Our experimental results show that the reduction cannot be fulfilled in the best form with standard time-series forecasting methods only (e.g., GRU) since the forecasting task is challenging, but \emph{our specific model design based on NCDEs enables the reduction}.
    \item{As exemplified in Fig.~\ref{fig:emb}, the base algorithms make a variety of latent dynamics, but our proposed method, TimeKit, still learns all kinds of the dynamics well.}
    \item Our experimental results show that our proposed TimeKit can be used for existing popular collaborative filtering algorithms and significantly improves them in all cases.
    \item{Traditional matrix factorization (MF)-based models may outperform advanced CF models (e.g. LightGCN) just by applying TimeKit, retaining their advantages --- lower time and space complexity.}
\end{compactenum}

\section{Related Work and Preliminaries}
In this section, we review recommender systems, dynamic embedding, and neural controlled differential equations (NCDEs).

\subsection{Recommender Systems}
\paragraph{\textbf{Collaborative Filtering}}
Traditional collaborative filtering recommender systems have focused on matrix factorization (MF) techniques~\cite{Koren09MF}. Typical MF-based methods include BPR~\cite{Rendle09BPR} and WRMF~\cite{Hu08WRMF}, and these MF-based methods simply learn relationships between users and items via dot-products. Therefore, they have limitations in considering potentially complex relationships between users and items inherent in user-item interactions~\cite{He17NeuMF}. To overcome these limitations, deep learning-based recommender systems, e.g., autoencoders~\cite{kingma2013auto,vincent2008extracting} and GCNs~\cite{Bruna2014,Defferrard2016,kipf2017semi,Wu2019SGC,Hamilton2017,Velickovic2018,Gao2018}, have been proposed to effectively learn more complicated relationships between users and items~\cite{Wang19NGCF,Ying18pinsage,Rianne17GCMC}.

Recently, recommender systems using GCNs~\cite{Wang19NGCF,Ying18pinsage,Rianne17GCMC} are gathering much attention. GCN-based methods can effectively learn the behavioral patterns between users and items by directly capturing the collaborative signals inherent in the user-item interactions~\cite{Wang19NGCF}. Typical GCN-based methods include GC-MC~\cite{Rianne17GCMC}, PinSage~\cite{Ying18pinsage}, and NGCF~\cite{Wang19NGCF}. In general, GCN-based methods model a set of user-item interactions as a user-item graph and perform the following three steps:

{\bfseries (Step 1) Initialization Step:} They randomly set the initial embedding $\bm{e}^0$ of all user $u$ and item $v$, denoted as follows:
\begin{equation}
    \begin{split}
    \bm  e_{u}^{0}, \bm e_{v}^{0} \in \mathbb {R}^{D},
    \end{split}
\end{equation}
where $D$ denotes the embedding size, $u \in U$ is a user, and $v \in V$ is an item.

\begin{table}
\footnotesize
\renewcommand*{\arraystretch}{0.95}
\center
  \caption{The characteristics of collaborative filtering recommender systems}
  
  \begin{tabular}{c|c|c|c|c}
  \toprule
  & BPRMF & NGCF & LightGCN & LT-OCF
\\ \midrule
Non-Linear Propagation & X & O & X & X \\
Linear Propagation & X & X & O & O \\
Residual Prediction & X & O & O & O \\
\bottomrule
\end{tabular}
\label{tab:GCNs}
\end{table}

{\bfseries (Step 2) Propagation Step:} First of all, this propagation step is iterated $K$ times, i.e., $K$ layers of embedding propagation. The embedding of a user node $u$ (resp. an item node $v$) in $i$-th layer is updated based on the embeddings of $u$'s (resp. $v$'s) neighbors $N_u$ (resp. $N_v$) in ($i-1$)-th layer as follows:

\begin{equation}\label{eq:prop}
    \begin{split}
    \bm  e_{u}^{i} = \sigma ( \Sigma_{v\in N_u} \bm e_v^{i-1} \bm{W}_{i} ), \;\;\;     \bm  e_{v}^{i} = \sigma ( \Sigma_{u\in N_v} \bm e_u^{i-1} \bm{W}_{i} ),
    \end{split}
\end{equation}
where $\sigma$ denotes a non-linear activation function, e.g., ~ReLU, and $\bm W_{i} \in \mathbb {R}^{D \times D} $ is a trainable transformation matrix. There also exist some other variations: i) including the self-embeddings, i.e., $N_u = N_u \cup \{u\}$ and $N_v = N_v \cup \{v\}$, ii) removing the transformation matrix, and/or iii) removing the non-linear activation, which is in particular called as \emph{linear propagation}~\cite{Chen20LRGCCF, He20LightGCN}.

{\bfseries (Step 3) Prediction Step:} The preference of user $u$ to item $v$ is predicted using the dot-product between the user $u$'s and item $v$'s embeddings in the last layer $K$, i.e., $\bm e_{u}^{K}$ and $\bm e_{v}^{K}$, as follows:
\begin{equation}\label{eq:pred}
    \begin{split}
    \hat{r}_{u,v} = \bm e_{u}^{K} \odot \bm e_v^K.
    \end{split}
\end{equation}

However, GCN-based methods have two limitations: i) training difficulty of using non-linear activation and ii) over-smoothing problem as the number of layers increases, i.e., too similar embeddings of nodes in the last layer~\cite{Wu2019SGC,chen2020measuring,Chen20LRGCCF,li2018deeper,He20LightGCN}.

{\bfseries (Step 3$'$) Alternative Prediction Step:} Recently, LR-GCCF~\cite{Chen20LRGCCF}, LightGCN~\cite{He20LightGCN}, and LT-OCF~\cite{choi2021ltocf}, which are GCN-based recommender systems to alleviate the problems, have been proposed. First, to alleviate the former problem, they remove non-linear activation functions. To mitigate the latter problem, they utilize the embeddings from all layers for prediction. After that, they perform \emph{residual prediction}~\cite{Chen20LRGCCF, He20LightGCN}, which predict each user's preference to each item with the multiple embeddings from the multiple layers, as follows:
\begin{equation}\label{eq:respred}
\setlength{\abovedisplayskip}{5pt}
\setlength{\belowdisplayskip}{5pt}
    \begin{split}
    \hat{r}_{u,v} = \bm e_{u}^{Last} \odot \bm e_v^{Last},
    \end{split}
\end{equation}where $\bm e_{u}^{Last} = \sum_{i=1}^K w_i\bm e_{u}^{i}$, $\bm e_{v}^{Last} = \sum_{i=1}^K w_i\bm e_{v}^{i}$, and $w_i$ is a coefficient.

In summary, we experiment with various models with different characteristics, and these can be characterized by, as shown in Table~\ref{tab:GCNs}, the propagation and prediction types. 


\begin{table}[t]
\centering
\renewcommand*{\arraystretch}{0.85}
\small
\caption{The comparison between the existing sequential recommendation and our proposed recommendation concepts}
\begin{tabular}{l|c|c}
\toprule
                                          & SR                                & TimeKit \\ \midrule
What does the model predict?              & Next items                        & \textbf{Embeddings}       \\
What is the input of model?               & \thead{Sequence of\\ items}       & \textbf{\thead{Sequence of\\ embeddings}}   \\ 
What type of prediction?                  & Point-wise                        & \textbf{Region-wise}   \\ 
\bottomrule
\end{tabular}
\label{tab:sr}
\end{table}


\paragraph{\textbf{Sequential Recommender Systems}}



We note that our TimeKit aims at a different task from the sequential recommendation as presented in Table~\ref{tab:sr}. The sequential recommendation is, given a sequence of items visited by a user, to predict one next item \cite{tang2018personalized, kang2018self, sun2019bert4rec}. While our main goal is to reduce the classical collaborative filtering to a time-series forecasting, the sequential recommendation typically makes a \emph{time point-wise} future item recommendation given a sequence of purchased items of a user. We forecast the future user/item embeddings, which allow us to predict items that a target user will buy during a certain future period, i.e., \emph{time region-wise} recommendation. Furthermore, the input is a sequence of embeddings (rather than a raw sequence of user-item interactions unlike the sequential recommendation).

\subsection{Dynamic Embedding}
One similar problem is \emph{dynamic embedding} where data arrives incrementally and we want to calculate their embedding vectors~\cite{DBLP:journals/corr/abs-1907-11968,DBLP:journals/corr/abs-1809-02657,egcn}. However, this sort of problem is different from our research since many of them have large model sizes for solving general downstream deep learning tasks. For instance, dynamic graph embedding methods consider the situation that graphs arrive incrementally and we want to embed those graphs over time into vectors to solve graph pattern classification problems. Another example is dynamic word embedding where we consider texts changing over time. On the contrary, our goal is to design a lightweight upgrade kit that can be readily integrated into existing collaborative filtering algorithms.

\subsection{Neural Controlled Differential Equations}
Neural controlled differential equations (NCDEs) are greatly inspired by neural ordinary differential equations (NODEs) and therefore, we first describe NODEs. NODEs solve the following integral problem to calculate $\bm{z}(T)$ from $\bm{z}(0)$~\cite{NIPS2018_7892}:
\begin{align}\label{eq:node}
    \bm{z}(T) = \bm{z}(0) + \int_{0}^{T}f(\bm{z}(t),t;\bm{\theta}_f)dt,
\end{align}where $f(\bm{z}(t),t;\bm{\theta}_f)$, which we call \emph{ODE function}, is a neural network to approximate $\dot{\bm{z}} \stackrel{\text{def}}{=} \frac{d \bm{z}(t)}{d t}$. To solve the integral problem, NODEs rely on ODE solvers, e.g., the explicit Euler method, the Dormand--Prince (DOPRI) method, and so forth~\cite{DORMAND198019}. We note that the integral problem in Eq.~\eqref{eq:node} uses the concept of Riemann integral.

Instead of the backpropagation method, the adjoint sensitivity method is used to train NODEs for its efficiency and theoretical correctness~\cite{NIPS2018_7892}. After letting $\bm{a}_{\bm{z}}(t) = \frac{d L}{d \bm{z}(t)}$ for a task-specific loss $L$, it calculates the gradient of loss w.r.t model parameters with another reverse-mode integral as follows:\begin{align}\nabla_{\bm{\theta}_f} L = \frac{d L}{d \bm{\theta}_f} = -\int_{T}^{0} \bm{a}_{\bm{z}}(t)^{\mathtt{T}} \frac{\partial f(\bm{z}(t), t;\bm{\theta}_f)}{\partial \bm{\theta}_f} dt.\end{align}

As shown in Eq.~\eqref{eq:ncde2}, on the other hand, NCDEs are written as $\bm{z}(T) = \bm{z}(0) + \int_{0}^{T} f(\bm{z}(t);\bm{\theta}_f) \frac{dX(t)}{dt} dt$,
where $X(t)$ is a continuous path created by an interpolation algorithm from a raw discrete time-series sample $\{(\bm{x}_i, t_i)\}_{i=0}^N$, where $t_i$ means the time-point of the observation $\bm{x}_i$, $t_0 = 0$, $t_N = T$ and $t_i < t_{i+1}$. Note that NCDEs keep reading the derivative of $X(t)$ over time, denoted $\dot{X}(t) \stackrel{\text{def}}{=} \frac{dX(t)}{dt}$. Therefore, NCDEs are considered as a continuous analogue to RNNs and are suitable for processing time-series data.

NCDEs use the Riemann--Stieltjes integral, as shown in Eq.~\eqref{eq:ncde2}. To solve it, existing ODE solvers can also be used since $\dot{\bm{z}}(t) \stackrel{\text{def}}{=} \frac{d\bm{z}(t)}{dt} = f(\bm{z}(t);\bm{\theta}_f) \frac{dX(t)}{dt}$ in NCDEs. Regardless of the integral problem type, existing ODE solvers can be used once $\dot{\bm{z}}(t)$ can be properly modeled and calculated. ODE solvers discretize time variable $t$ and convert an integral into many steps of additions~\cite{10.2307/j.ctvzsmfgn}, i.e., update $\bm{z}(t + s)$ from $\bm{z}(t)$. For instance, the fourth-order Runge--Kutta (RK4) method uses the following method:
\begin{align}\label{eq:rk4}
\bm{z}(t + s) = \bm{z}(t) + \frac{s}{6}\Big(f_1 + 2f_2 + 2f_3 + f_4\Big),
\end{align}where $s$, which is usually smaller than 1, is a pre-determined step size, $f_1 = f(\bm{z}(t), t;\bm{\theta}_f)$, $f_2 = f(\bm{z}(t) + \frac{s}{2}f_1, t+\frac{s}{2};\bm{\theta}_f)$, $f_3 = f(\bm{z}(t) + \frac{s}{2}f_2, t+\frac{s}{2};\bm{\theta}_f)$, and $f_4 = f(\bm{z}(t)+sf_3, t+s;\bm{\theta}_f)$.

The Dormand-Prince(DOPRI) method is one of the most advanced solvers, which decides the step-size $s$ every step on its own. In other words, it is an $adaptive$ step-size solver. NODEs and NCDEs use RK4 or DOPRI in many cases.

\section{Proposed Method}
We describe our method in detail in this section. We first explain the overall workflow in our method, followed by detailed model and training method designs.

\begin{figure}[t]
    \centering
    \includegraphics[width=\columnwidth]{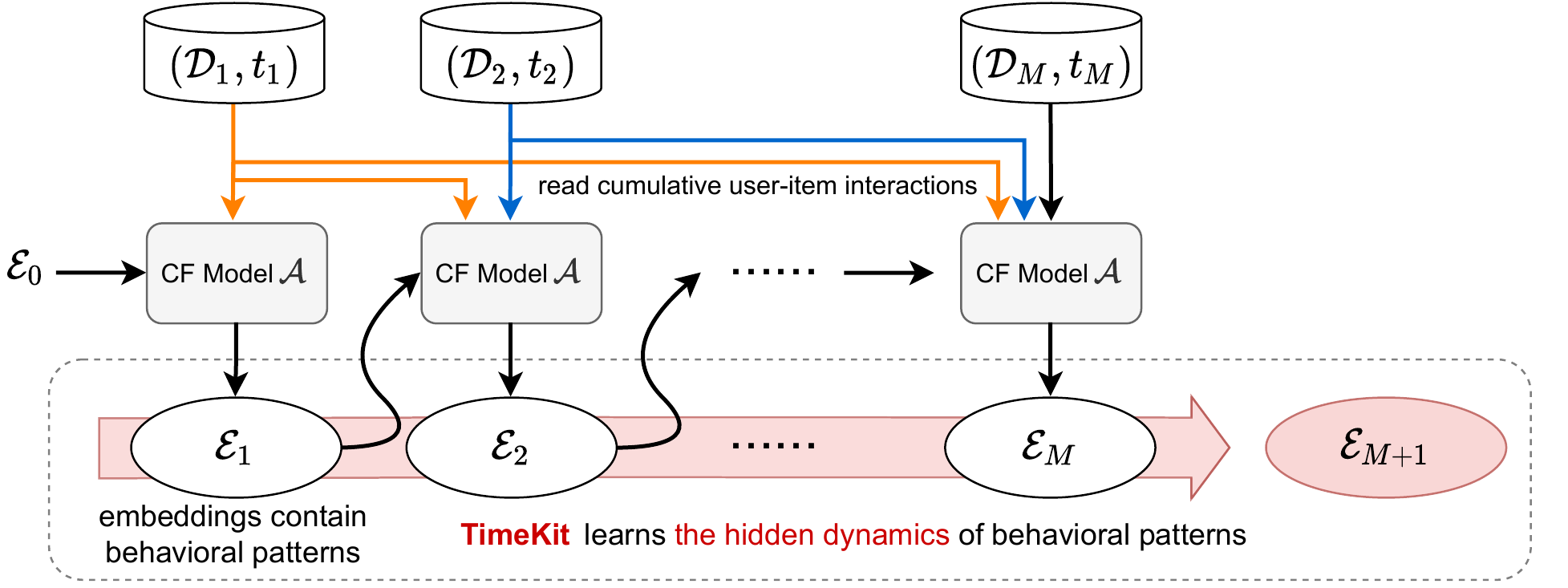}
    \caption{The overall workflow of creating the time-series of embedding vectors and forecasting. We do not train a recommendation model, e.g., LightGCN, from scratch every period, but incrementally after reusing the most recently trained parameters, e.g., initial embeddings in the case of LightGCN. We use all past data on or before each $t_{i}$ to avoid overfitting on new items.}
    \label{fig:create}
\end{figure}


\subsection{Overall Workflow}
The detailed workflow in our method to forecast future embedding vectors is as follows --- Fig.~\ref{fig:create} shows how to create a time-series of embedding vectors and how we forecast:
\begin{compactenum}
    \item Let $U$ (resp. $V$) be a set of users (resp. items). Given a long history of user-item interactions and its time span $[0,T]$, we define a set of data, denoted $\{(\mathcal{D}_i, t_i)\}_{i=1}^{M+1}$, where $\mathcal{D}_i$ is a set of interactions during the period of $(t_{i-1},t_{i}]$ and $t_M$ (resp. $t_{M+1}$) is the last time-point of our training/validating (resp. testing) data. We note that $t_{i-1} < t_i$, $t_0 = 0$, and $t_{M+1} = T$. At $t_0$, there are no interactions yet.
    \item Starting from $(\mathcal{D}_1, t_1)$, we incrementally train a recommendation algorithm $\mathcal{A}$ which produces user and item embedding vectors. In most algorithms, user and item embedding vectors are the only trainable parameters, e.g., the initial embeddings $\{\bm  e_{u}^{0}\}_{u=1}^{|U|}, \{\bm e_{v}^{0}\}_{v=1}^{|V|}$ of LightGCN, where no other parameters are trained. When training for $(\mathcal{D}_i, t_i)$, we initialize user and item embedding vectors with those trained for $(\mathcal{D}_{i-1}, t_{i-1})$, and further train with $\cup_{j=1}^{i} \mathcal{D}_j$ --- in other words, we consider all the previous interactions on or before $t_i$ and \emph{incrementally} train the parameters while maintaining a single vector space across $[0,T]$. Since we do not randomly initialize user and item embedding vectors every time but initialize with the previously trained one, one embedding vector space can be maintained. Moreover, it can bring benefits, e.g., training efficiency, because the initialized parameters are already somehow good.
    \item Let $\{(\mathcal{E}_i, t_i)\}_{i=1}^M$ be a time-series of embedding vectors produced in the above way, e.g., $\mathcal{E}_i=\{\bm e_{u}^{Last}\}_{u=1}^{|U|} \cup \{\bm e_{v}^{Last}\}_{v=1}^{|V|}$ in the case of LightGCN (trained up to $t_i$) as in Eq.~\eqref{eq:respred}, or $\mathcal{E}_i=\{\bm e_{u}^{K}\}_{u=1}^{|U|} \cup \{\bm e_{v}^{K}\}_{v=1}^{|V|}$ if residual prediction is not used as in Eq.~\eqref{eq:pred}. Using our time-series forecasting method, we forecast future embeddings, $\hat{\mathcal{E}}_{M+1}$ --- we use the symbol of ``$\hat{\;}$'' to denote predictions.
    \item Using $\hat{\mathcal{E}}_{M+1}$, we recommend items to users after calculating their dot-products.
\end{compactenum}


\subsection{Rationale behind Our Method} In real-world applications, we need to train a collaborative filtering algorithm $\mathcal{A}$ incrementally since new user-item interactions are accumulated everyday. One typical method is to retrain $\mathcal{A}$ and replace the previous model with the retrained model --- one can also use only recent data when retraining $\mathcal{A}$ in order to reflect recent patterns. However, this typical use scenario does not fully exploit the time-series characteristic of the user and item embedding vectors over time.

\begin{figure}[t]
    \centering
    \subfigure[User 13177's embedding (NGCF)]{\includegraphics[width=0.49\columnwidth]{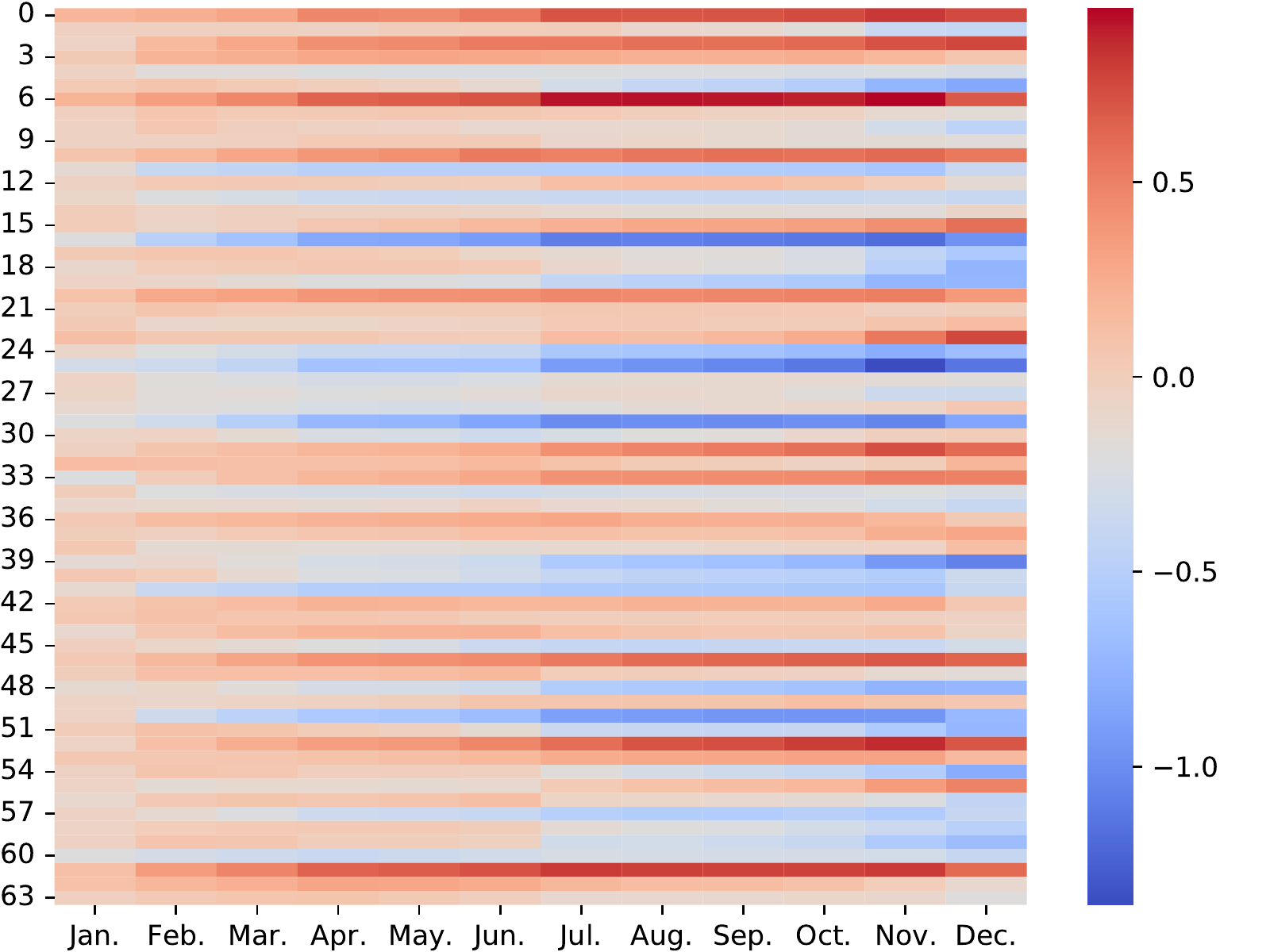}}
    \subfigure[Item 2240's embedding (NGCF)]{\includegraphics[width=0.49\columnwidth]{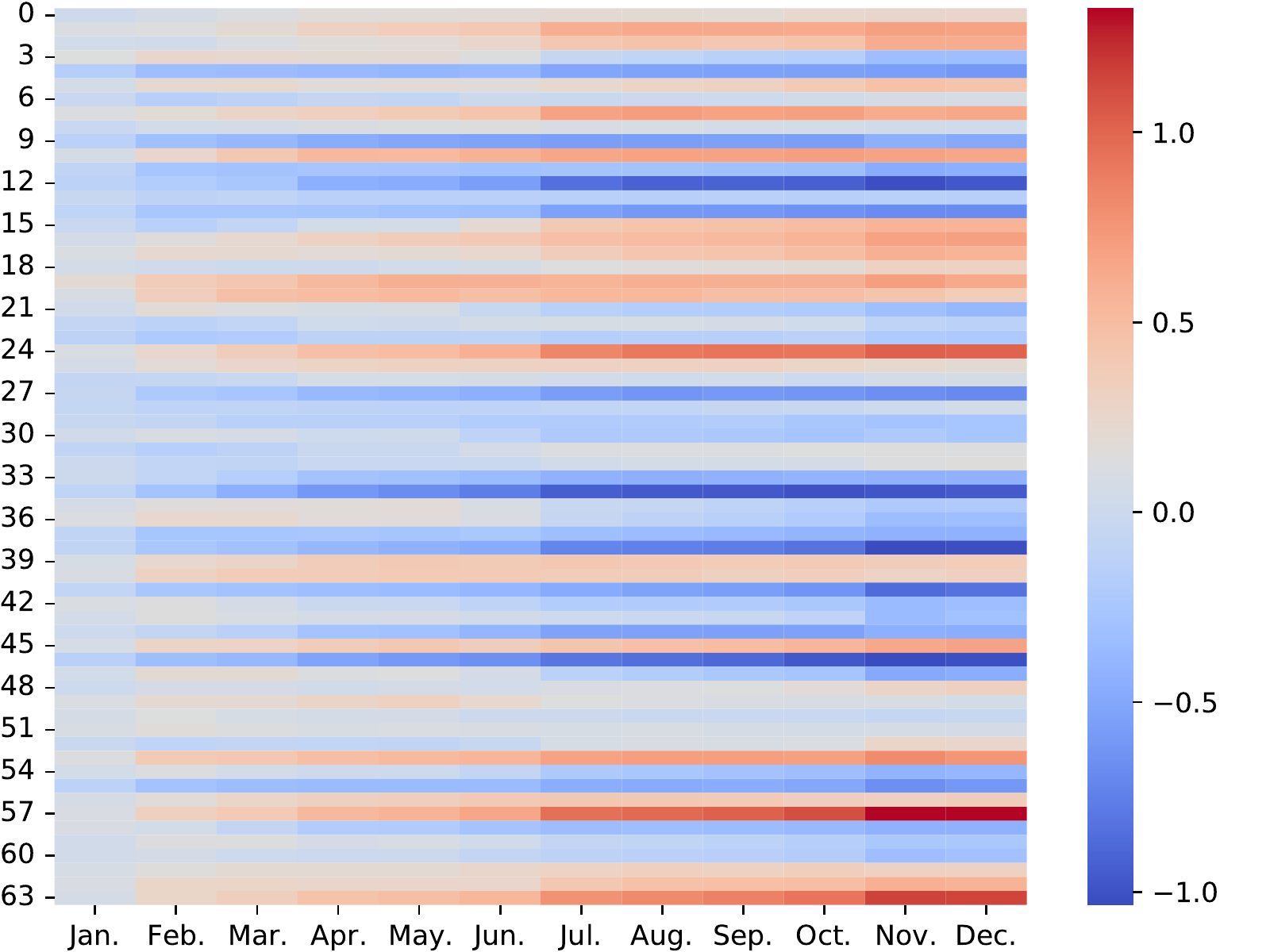}}
    \caption{Visualization of 64-dimensional user/item embedding vector changes over 12 months. We note that the embeddings of December (in the rightmost position) are the ones forecast by TimeKit on top of NGCF in Netflix. Note that the embedding vectors change gradually over time, but in different patterns depending on base algorithms. Month is on the X-axis, while the index of the embedding vector is on the Y-axis.} 
    \label{fig:emb}
\end{figure}

In this work, we are not interested in designing such retraining-replacement methods but in forecasting \emph{future} embedding vectors after capturing the hidden dynamics describing the evolutionary process of the embedding vectors. Psychologically, people's preferences on items are affected by various factors and fluctuate a lot over time. Therefore, understanding the user behavioral pattern drift\footnote{The user behavioral pattern drift means that a user $u$'s preference on an item $v$ given a user-item interaction data $\mathcal{D}_i$ collected by time $t_i$ is different from that given another data $\mathcal{D}_j$ collected by $t_j$, i.e., $Pr(v|u,D_i) \neq Pr(v|u,D_j)$ if $i \neq j$.} can be a key in real-world collaborative filtering applications. We conjecture that such hidden dynamics that defines the drift exists and our method is able to learn. As shown in Fig.~\ref{fig:emb}, we found that those embedding vectors do not change randomly over time but are likely to change following a hidden dynamics. Therefore, we conjecture that by revealing the hidden dynamics, we may be able to better perform the collaborative filtering task.

However, we found that forecasting future embedding vectors is a challenging extrapolation task. In particular, vanilla RNNs are insufficient for learning the complicated temporal patterns of the item embedding vectors. We also found that recent differential equation-based breakthroughs in processing complicated time-series data can enable our proposed recommendation concept.

A differential equation (or a governing equation) means an equation describing $\dot{\bm{z}} \stackrel{\text{def}}{=} \frac{d \bm{z}(t)}{d t}$ at any time $t$, e.g, the Navier–Stokes equation describing fluid dynamics~\cite{doi:10.2514/1.J052606}, the Kermack-McKendrick equation describing infectious disease dynamics~\cite{SIR}, and so on. Once a correct equation exists and can be learned, the time-evolving process of $\bm{z}(t)$ can be reproduced regardless of how far it does extrapolate. We attempt to approximate such a human behavioral dynamics for recommendation with NCDEs.

\subsection{Embedding Forecaster}
Our forecasting task is basically a many-to-one forecasting task, i.e., reading multiple recent embeddings for forecasting the very next embedding. Our task can be formally described as forecasting $\hat{\mathcal{E}}_{i+1}$ after reading $\{(\mathcal{E}_{i-r}, t_{i-r})\}_{r=0}^{R-1}$ at a certain time-point $t_i$, i.e., reading recent $R$ periods to forecast for the very next period.

Let $\{(\bm{x}_i, t_i)\}_{i=1}^M$ be a time-series of embedding vectors for a user (or an item). For simplicity but without loss of generality, we describe how to forecast $\hat{\bm{x}}_{i+1}$ from $\{(\bm{x}_{i-r}, t_{i-r})\}_{r=0}^{R-1}$ since we apply the same forecasting method to each user and item --- we note that we then can create $M-R$ training samples for each user (or each item). To this end, we design GRU and NCDE-based time-series forecasting models.

\paragraph{\textbf{GRU-based model}}
From $\{(\bm{x}_{i-r}, t_{i-r})\}_{r=0}^{R-1}$, we can use GRUs to predict $\hat{\bm{x}}_{i+1}$, which is a typical many-to-one forecasting task. In comparison with LSTMs, GRUs provide as high model capacity as that of LSTM but have a lower complexity~\cite{chung2014empirical}. The forecasting with GRUs can be written as follows:
\begin{align}\begin{split}\label{eq:gru}
\bm{h}_{i-R} &= \bm{0},\\
\bm{h}_j &= GRU(\bm{h}_{j-1}, \bm{x}_{j}), i-R+1 \leq j \leq i\\
\hat{\bm{x}}_{i+1} &= \Phi_{GRU}(\bm{h}_{i}; \bm{\theta}_{\Phi_{GRU}}),
\end{split}\end{align}where $\Phi_{GRU}$ is an output fully-connected layer parameterized by $\bm{\theta}_{\Phi_{GRU}}$, and $\bm{h}$ is a hidden vector.

\paragraph{\textbf{NCDE-based model}}
We also use the following NCDE-based method to forecast the future embedding vectors:
\begin{align}\begin{split}\label{eq:ourncde}
\bm{z}(t_{i-R+1}) &= \xi_{CDE}(\bm{x}_{i-R+1}; \bm{\theta}_{\xi_{CDE}}),\\
\bm{z}(t_{i}) &= \bm{z}(t_{i-R+1}) + \int_{t_{i-R+1}}^{t_{i}} f(\bm{z}(t);\bm{\theta}_f) \frac{dX(t)}{dt} dt,\\
\hat{\bm{x}}_{i+1} &= \Phi_{CDE}(\bm{z}(t_{i}); \bm{\theta}_{\Phi_{CDE}}),
\end{split}\end{align}where $X$ is a continuous path in $[t_{i-R+1}, t_{i}]$ created from $\{(\bm{x}_{i-r}, t_{i-r})\}_{r=0}^{R-1}$ by the natural cubic spline interpolation algorithm, $\xi_{CDE}$ is an initial value generation layer, $\Phi_{CDE}$ is an output layer. We use fully connected-layers to define $\xi_{CDE}$ and $\Phi_{CDE}$. The path created by the natural cubic spline is smooth and twice differentiable, which are required to calculate the gradients w.r.t. $\bm{\theta}_f$. Therefore, it shows the best fit to NCDEs among available interpolation algorithms. The key point is the following definition of the CDE function $f$:
\begin{align}
\sigma(FC_2(\sigma(FC_1(\bm{z}(t)))))
\end{align}where $FC_i$ means the $i$-th fully connected layer and $\sigma$ is a non-linear activation, such as hyperbolic tangent. We intentionally choose these operators to define $f$ since they are all Lipschitz-continuous, which guarantees the following well-posedness of training $\bm{\theta}_f$, whereas one can choose other advanced definitions for $f$~\cite{xia2021heavy}. The well-posedness\footnote{A well-posed problem means i) its solution uniquely exists, and ii) its solution continuously changes as input data changes.} of NCDEs was already proved in \cite[Theorem 1.3]{lyons2004differential} under the mild condition of the Lipschitz continuity. Almost all activations, such as ReLU, ELU, Leaky ReLU, SoftPlus, Tanh, Sigmoid, ArcTan, and Softsign, have a Lipschitz constant of 1. Other common neural network layers, such as dropout, batch normalization and other pooling methods, have explicit Lipschitz constant values. Therefore, the Lipschitz continuity of $f$ can be fulfilled in our case. This makes our training problem for NCDEs well-posed. As a result, our training algorithm solves a well-posed problem so its training process is stable in practice.

\subsection{Training Algorithm}\label{sec:tr}
We first note that we maintain two forecasting models: one for user embedding vectors and the other for item embedding vectors. Overall, our training method follows the standard multivariate time-series forecasting model training process - minimizing the mean squared error (MSE) loss. Given a set of users (resp. items), we create a mini-batch of $B$ users (resp. items) and train our model for them every iteration. We use the Adam optimizer with a learning rate $\lambda$. We use a validation set to update the best model.

\begin{table}
\setlength{\tabcolsep}{4pt}
\renewcommand*{\arraystretch}{0.9}
\centering
\small
\caption{Statistics of datasets}
\label{t3}
\begin{tabular}{cccccccc}
\toprule
& User \# & Item \# & Interaction \# & Density \\
\midrule 
Gowalla & 2,970 & 69,853 & 609,582 & 0.00293 \\
Amazon-Book & 18,654 & 49,972 & 872,135 & 0.00094 \\
Goodreads & 18,576 & 109,727 & 1,142,707 & 0.00056 \\
Netflix & 22,060 & 17,059 & 3,783,546 & 0.01005 \\

\bottomrule
\end{tabular}
\end{table}

\begin{table*}[t]
\caption{The best hyperparameters of our forecasting models}
\scriptsize
\centering
\renewcommand*{\arraystretch}{0.85}
\setlength{\tabcolsep}{3pt}
\label{tab:hyperparam}
\begin{tabular}{c|c|cccc|cccc|cccc}
\toprule
\multirow{4}{*}{Dataset} & \multirow{4}{*}{Base Alg. $\mathcal{A}$} & \multicolumn{4}{c|}{GRU-GRU}     & \multicolumn{4}{c|}{NCDE-NCDE}          & \multicolumn{4}{c}{GRU-NCDE}    \\ \cmidrule{3-14} 
                         & & \multicolumn{2}{c|}{User Embedding} & \multicolumn{2}{c|}{Item Embedding} & \multicolumn{2}{c|}{User Embedding}     & \multicolumn{2}{c|}{Item Embedding} & \multicolumn{2}{c|}{User Embedding}     & \multicolumn{2}{c}{Item Embedding} \\ \cmidrule{3-14} 
                         & & hidden & \multicolumn{1}{c|}{$\lambda$} & hidden     & $\lambda$    & hidden & \multicolumn{1}{c|}{$\lambda$} & hidden     & $\lambda$           & hidden & \multicolumn{1}{c|}{$\lambda$} & hidden           & $\lambda$          \\ \midrule
\multirow{4}{*}{Gowalla}     & BPRMF                   &     256   & \multicolumn{1}{c|}{$5.0 \times e^{-6}$}          &           64              &         $1.0 \times e^{-5}$     &    256    & \multicolumn{1}{c|}{$1.0 \times e^{-3}$}          &          128        &               $1.0 \times e^{-3}$      &     256   & \multicolumn{1}{c|}{$5.0 \times e^{-6}$}          &       256           &           $5.0 \times e^{-4}$         \\ \cline{2-14} 
                             & NGCF                    &    128    & \multicolumn{1}{c|}{$1.0 \times e^{-5}$}          &                512         &        $5.0 \times e^{-6}$      &    64    & \multicolumn{1}{c|}{$1.0 \times e^{-3}$}          &  128                &           $5.0 \times e^{-4}$          &    512    & \multicolumn{1}{c|}{$1.0 \times e^{-4}$}          &        64          &         $1.0 \times e^{-3}$           \\ \cline{2-14} 
                             & LightGCN                &  512    & \multicolumn{1}{c|}{$1.0 \times e^{-5}$}          &      128            &     $1.0 \times e^{-5}$        &    128    & \multicolumn{1}{c|}{$5.0 \times e^{-3}$}          &      64           &     $5.0 \times e^{-4}$                &    128    & \multicolumn{1}{c|}{$5.0 \times e^{-5}$}          &   64               &            $5.0 \times e^{-4}$        \\ \cline{2-14} 
                             & LT-OCF                 &   128     & \multicolumn{1}{c|}{$1.0 \times e^{-5}$}          &              64           &       $1.0 \times e^{-4}$       &   256     & \multicolumn{1}{c|}{ $1.0 \times e^{-3}$}          &        256      &           $1.0 \times e^{-5}$        &   512     & \multicolumn{1}{c|}{$1.0 \times e^{-5}$}          &       256           &       $1.0 \times e^{-5}$                  \\ \hhline{==============}
\multirow{4}{*}{Amazon-book} & BPRMF                   &    512    & \multicolumn{1}{c|}{$5.0 \times e^{-5}$}          &       64                  &       $5.0 \times e^{-6}$       &    128    & \multicolumn{1}{c|}{$1.0 \times e^{-3}$}          &         256         &             $5.0 \times e^{-4}$        &    128    & \multicolumn{1}{c|}{$5.0 \times e^{-5}$}          &        256          &          $5.0 \times e^{-4}$          \\ \cline{2-14} 
                             & NGCF                    &     128   & \multicolumn{1}{c|}{$5.0 \times e^{-6}$}          &           64              &     $5.0 \times e^{-5}$         &     128   & \multicolumn{1}{c|}{$1.0 \times e^{-3}$}          &   256               &         $5.0 \times e^{-4}$            &    256    & \multicolumn{1}{c|}{$1.0 \times e^{-5}$}          &        256          &            $5.0 \times e^{-4}$        \\ \cline{2-14} 
                             & LightGCN                &    64    & \multicolumn{1}{c|}{$5.0 \times e^{-5}$}          &        64                 &      $1.0 \times e^{-5}$        &    128    & \multicolumn{1}{c|}{$1.0 \times e^{-3}$}          &         256         &      $5.0 \times e^{-4}$               &   512     & \multicolumn{1}{c|}{$5.0 \times e^{-5}$}          &     256             &            $5.0 \times e^{-4}$        \\ \cline{2-14} 
                             & LT-OCF                  &   128     & \multicolumn{1}{c|}{$1.0 \times e^{-6}$}          &           256              &        $1.0 \times e^{-4}$      &    128    & \multicolumn{1}{c|}{$1.0 \times e^{-3}$}          &         256         &   $5.0 \times e^{-4}$                  &    512    & \multicolumn{1}{c|}{$1.0 \times e^{-4}$}          &         256         &        $5.0 \times e^{-4}$            \\ \hhline{==============}
\multirow{4}{*}{Goodreads}   & BPRMF                   &   256     & \multicolumn{1}{c|}{$1.0 \times e^{-6}$}          &        256                &     $1.0 \times e^{-6}$         &    256    & \multicolumn{1}{c|}{$1.0 \times e^{-3}$}          &       128           &               $5.0 \times e^{-4}$      &    256    & \multicolumn{1}{c|}{$1.0 \times e^{-6}$}          &          128        &          $5.0 \times e^{-4}$         \\ \cline{2-14} 
                             & NGCF                    &   64     & \multicolumn{1}{c|}{$1.0 \times e^{-6}$}          &              64           &       $1.0 \times e^{-6}$       &     64   & \multicolumn{1}{c|}{$1.0 \times e^{-3}$}          &          64        &         $5.0 \times e^{-6}$            &   64    & \multicolumn{1}{c|}{$1.0 \times e^{-6}$}          &        64          &      $5.0 \times e^{-6}$              \\ \cline{2-14} 
                             & LightGCN                &    256    & \multicolumn{1}{c|}{$5.0 \times e^{-5}$}          &   64                      &      $5.0 \times e^{-6}$        &     128   & \multicolumn{1}{c|}{$5.0 \times e^{-4}$}          &        256          &     $5.0 \times e^{-4}$                &    512    & \multicolumn{1}{c|}{$5.0 \times e^{-6}$}          &          256        &          $5.0 \times e^{-4}$          \\ \cline{2-14} 
                             & LT-OCF                  &    128    & \multicolumn{1}{c|}{$1.0 \times e^{-5}$}          &          64               &     $1.0 \times e^{-5}$         &     128   & \multicolumn{1}{c|}{$1.0 \times e^{-3}$}          &     256            &         $5.0 \times e^{-4}$            &   512    & \multicolumn{1}{c|}{$5.0 \times e^{-6}$}          &        256          &              $5.0 \times e^{-4}$     \\  \hhline{==============}
\multirow{4}{*}{Netflix}     & BPRMF                   &    64    & \multicolumn{1}{c|}{$1.0 \times e^{-6}$}          &           128              &      $5.0 \times e^{-5}$        &    128    & \multicolumn{1}{c|}{$5.0 \times e^{-4}$}          &        256          &    $1.0 \times e^{-3}$                 &     64   & \multicolumn{1}{c|}{$1.0 \times e^{-6}$}          &            256      &         $1.0 \times e^{-3}$           \\ \cline{2-14} 
                             & NGCF                    &     512   & \multicolumn{1}{c|}{$5.0 \times e^{-5}$}          &             256            &        $5.0 \times e^{-6}$      &   256     & \multicolumn{1}{c|}{$1.0 \times e^{-3}$}          &   256               &         $1.0 \times e^{-4}$            &    256    & \multicolumn{1}{c|}{$5.0 \times e^{-6}$}          &         256         &        $1.0 \times e^{-4}$            \\ \cline{2-14} 
                             & LightGCN                &    64    & \multicolumn{1}{c|}{$1.0 \times e^{-3}$}          &         256                &       $5.0 \times e^{-5}$       &    256    & \multicolumn{1}{c|}{$5.0 \times e^{-5}$}          &    64              &    $5.0 \times e^{-4}$                 &    128    & \multicolumn{1}{c|}{$1.0 \times e^{-6}$}          &           64       &             $5.0 \times e^{-4}$       \\ \cline{2-14} 
                             & LT-OCF                  &    64    & \multicolumn{1}{l|}{$5.0 \times e^{-6}$}          & \multicolumn{1}{c}{64}    &       $5.0 \times e^{-5}$       &    128    & \multicolumn{1}{c|}{$5.0 \times e^{-4}$}          &         64         &    $5.0 \times e^{-4}$           &    128    & \multicolumn{1}{c|}{$5.0 \times e^{-5}$}          &         64         &        $1.0 \times e^{-3}$            \\ \bottomrule
\end{tabular}
\end{table*}

\begin{table*}[t]
\setlength{\tabcolsep}{3pt}
\renewcommand*{\arraystretch}{0.95}
\centering
\scriptsize
\caption{Main experimental results with four base collaborative filtering algorithms, where the best results among the base algorithms are in gray, the best results among all methods are in boldface and the best improvement ratios are underlined.} 
\label{tab:main_score}
\begin{tabular}{c|c|cc|cc|cc|cc|cc}
\toprule
\multirow{3}{*}{Dataset}    & \multirow{3}{*}{Base Alg. $\mathcal{A}$} & \multicolumn{2}{c|}{Original} & \multicolumn{2}{c|}{GRU-GRU} & \multicolumn{2}{c|}{NCDE-NCDE} & \multicolumn{2}{c|}{GRU-NCDE} & \multicolumn{2}{c}{Improvement (\%)} \\ \cmidrule{3-12} 
                            &                   & Recall@20       & NDCG@20      & Recall@20      & NDCG@20     & Recall@20      & NDCG@20     & Recall@20      & NDCG@20     & Recall@20         & NDCG@20         \\ \midrule
\multirow{4}{*}{Gowalla}
                            & BPRMF             & 0.0292          & 0.0210       & \textbf{0.0323}   & 0.0222      & 0.0296         & 0.0215      & 0.0320         & \textbf{0.0227}    & 10.69             & 8.34            \\ \cline{2-12} 
                            & NGCF              & 0.0285          & 0.0198       &\textbf{0.0289}        & 0.0202      & 0.0285         & \textbf{0.0205}     & 0.0280         & 0.0204      & 1.60              & 3.50            \\ \cline{2-12} 
                            & LightGCN          & 0.0302          & 0.0236       & \textbf{0.0322}         & 0.0237      & 0.0317         & 0.0236      & 0.0312         & \textbf{0.0238}      & 6.66              & 0.66            \\ \cline{2-12} 
                            & LT-OCF            & \cellcolor[HTML]{BCBCBC}0.0345          & \cellcolor[HTML]{BCBCBC}0.0263       & \textbf{0.0454}         & \textbf{0.0363}      & 0.0329         & 0.0238      & 0.0324         & 0.0248      & \underline{31.57}             &\underline{37.86}     \\ \hhline{============}
\multirow{4}{*}{Amazon-Book}
                            & BPRMF             & 0.0496          & 0.0253       & 0.0546         & 0.0281      & 0.0571         & 0.0300      & \textbf{0.0575}         & \textbf{0.0303}      & 15.88             & 19.50           \\ \cline{2-12} 
                            & NGCF              & 0.0359          & 0.0181       & 0.0386         & 0.0193      & 0.0470         & 0.0245      & \textbf{0.0471}         & \textbf{0.0252}      & \underline{31.12}             & \underline{39.58}           \\ \cline{2-12} 
                            & LightGCN          & 0.0565          & 0.0294       & 0.0597         & 0.0313      & 0.0654         & 0.0355      & \textbf{0.0667}         & \textbf{0.0365}      & 17.93             & 24.17           \\ \cline{2-12} 
                            & LT-OCF            & \cellcolor[HTML]{BCBCBC}0.0573          & \cellcolor[HTML]{BCBCBC}0.0299       & 0.0674         & 0.0361      & 0.0681         & 0.0370      & \textbf{0.0711}         & \textbf{0.0393}      & 24.11             & 31.40           \\\hhline{============}
\multirow{4}{*}{Goodreads}
                            & BPRMF             & 0.0298          & 0.0171       & 0.0330         & 0.0189      & 0.0320         & 0.0184      & \textbf{0.0337}         & \textbf{0.0192}      & 13.04             & 12.21           \\ \cline{2-12} 
                            & NGCF              & 0.0235          & 0.0129       & 0.0242         & 0.0136      & 0.0270         & 0.0174      & \textbf{0.0292}         & \textbf{0.0191}      & \underline{23.97}             & \underline{47.69}           \\ \cline{2-12} 
                            & LightGCN          & \cellcolor[HTML]{BCBCBC}0.0359          & \cellcolor[HTML]{BCBCBC}0.0205       & 0.0365         & 0.0210      & 0.0364         & 0.0208      & \textbf{0.0377}         & \textbf{0.0214}      & 4.85              & 4.36            \\\cline{2-12}
                            & LT-OCF            & 0.0358          & 0.0205       & 0.0362         & 0.0207      & 0.0385         & 0.0220      & \textbf{0.0389}         & \textbf{0.0221}      & 8.64              & 8.04            \\ \hhline{============}
\multirow{4}{*}{Netflix}
                            & BPRMF             & 0.0701          & 0.0405       & 0.0713         & 0.0416      & 0.0774         & 0.0462      & \textbf{0.0776}         & \textbf{0.0466}      & 10.74             & 15.07           \\ \cline{2-12} 
                            & NGCF              & 0.0608          & 0.0380       & 0.0777         & 0.0515      & 0.1023         & 0.0641      & \textbf{0.1024}         & \textbf{0.0642}      & \underline{68.46}             & \underline{68.75}           \\ \cline{2-12} 
                            & LightGCN          & \cellcolor[HTML]{BCBCBC}0.0787          & \cellcolor[HTML]{BCBCBC}0.0451       & 0.0816         & 0.0471      & 0.0813         & 0.0468      & \textbf{0.0823}         & \textbf{0.0480}      & 4.52              & 6.33            \\\cline{2-12}
                            & LT-OCF            & 0.0779          & 0.0446       & 0.0804         & 0.0463      & 0.0859         & 0.0491      & \textbf{0.0899}         & \textbf{0.0539}      & 15.41              & 20.92            \\ 
\bottomrule
\end{tabular}
\end{table*}

\section{Experiments}
\subsection{Experiments Environments}
\subsubsection{Datasets}
For evaluation, we used the following four real-world datasets that are all publicly available to download:
\begin{itemize}
    \item {Gowalla's data is from January to October of 2010.}
    \item {Amazon-Book is a book rating data from 2015 to 2016 of Amazon-book dataset.}
    \item Goodreads contains book reviews from the Goodreads. The data for the two years 2015 and 2016 is used.
    \item Netflix contains user-movie monthly interactions of 2005.
    \item Table~\ref{t3} shows the detailed statistics of four datasets.
\end{itemize}


\subsubsection{Base Recommendation Algorithms} \label{sec:base_recommendation} Our proposed method does not assume any specific choice on the underlying recommendation algorithm. However, we choose to test with i) the conventional matrix factorization with the Bayesian personalized ranking loss (BPRMF), ii) NGCF, iii) LightGCN, and iv) LT-OCF\footnote{BPRMF and LightGCN are from https://github.com/gusye1234/LightGCN-PyTorch. NGCF is from https://github.com/huangtinglin/NGCF-PyTorch, and LT-OCF is from https://github.com/jeongwhanchoi/LT-OCF.}. BPRMF is one of the most classical collaborative filtering methods and is still widely used for many services, NGCF is a representative non-linear propagation-based method, and LightGCN and LT-OCF well represent recent research trends in graph-based collaborative filtering. We have chosen this specific set of base algorithms considering their diverse characteristics.

\subsubsection{Time-series Forecasting Methods}
We consider the following time-series forecasting models:
\begin{itemize}
    \item GRUs in Eq.~\eqref{eq:gru} are one of the most widely used RNN types. It provides more efficient computation with comparable capabilities in comparison with LSTMs.
    \item NCDEs in Eq.~\eqref{eq:ourncde} are a breakthrough in processing complicated dynamics. Among many differential equation-based time-series forecasting technologies, it shows the state-of-the-art performance in many real-world applications~\cite{jhin2021ancde,choi2022STGNCDE}.
\end{itemize}

Our method is marked with the naming convention of ``[User embedding forecasting method]-[Item embedding forecasting method],'' e.g., GRU-NCDE means that we use GRU in Eq.~\eqref{eq:gru} for forecasting user embeddings, and the NCDE design in Eq.~\eqref{eq:ourncde} for forecasting item embeddings.

\subsubsection{Evaluation Methods}
For all datasets, we split them into training/validating/testing sets. Setting a suitable period is crucial for the training of embedding vectors because a sufficient number of interactions must be guaranteed to find behavior dynamics. We set the period to a month for Gowalla and Netflix, which items are consumed frequently. Because books are not commonly consumed, for Amazon-book and Goodreads, we merge three months into one period to ensure that a sufficient number of interactions is guaranteed. In real-word settings, there will be a model update cycle depending on the domain, and a period can be from one model update to next. The details of constructing data are as follows:
\begin{itemize}
    \item For Gowalla, 10 time periods are produced from ten months of data, as we use a month as one period.
    \item For Amazon-book and Goodreads, we use two years of data into 8 periods by combining three months.
    \item For Netflix, each period data consists of interactions in each month. One year of data is split into 12 periods.
\end{itemize}

We first train and validate the base collaborative filtering algorithms with the user-item interaction logs in $\{(\mathcal{D}_i, t_i)\}_{i=1}^{M}$ and recommend for $(t_M, t_{M+1}]$, denoted as ``Original.'' Then, we upgrade them with time-series forecasting methods, denoted with each forecasting method, i.e., GRU-GRU, NCDE-NCDE, and GRU-NCDE. We train and validate the upgraded method in the seq-to-seq forecasting fashion. During the testing process, therefore, those upgraded methods read $\{(\mathcal{E}_i, t_i)\}_{i={M-R+1}}^M$ and forecast $\mathcal{E}_{M+1}$ with which we perform the collaborative filtering task. 

\subsubsection{Environmental Setting}
Our software and hardware environments are as follows: \textsc{Ubuntu} 18.04 LTS, \textsc{Python} 3.7.10, \textsc{Torch} 1.9.0, \textsc{CUDA} 10.0, and \textsc{NVIDIA} Driver 450.102.04, i9 CPU, and \textsc{NVIDIA RTX Titan}. The recommended ranges and the best configurations are as follows: 

\paragraph{\textbf{Preprocessing to prepare embedding vectors}} The learning rate is set to $1.0 \times e^{-3}$ and train 500 epochs with a mini-batch size of 1024 or 2048 for all but one case --- we use learning rate of $1.0 \times e^{-4}$ when training Netflix using NGCF. In the case of NGCF, the weight decay is set to $1.0 \times e^{-4}$ for Gowalla, Amazon-book, and Goodreads, and $1.0 \times e^{-5}$ for Netflix. In other models, for Gowalla and Netflix, the optimal weight decay is $1.0 \times e^{-3}$, whereas for other datasets, $1.0 \times e^{-4}$. We set the number of layers to 3 for training NGCF, LightGCN and LT-OCF, i.e., $K=3$ in Eq.~\eqref{eq:respred}.

\paragraph{\textbf{Forecasting}} The best hyperparameters of our forecasting models are shown in Table~\ref{tab:hyperparam}. We train GRU and NCDEs with a hidden size of \{64, 128, 256, 512\} and a weight decay of \{$1.0 \times e^{-4}$, $1.0 \times e^{-5}$\}. The learning rate is set over a range of intervals --- \{$5.0 \times e^{-3}$, $1.0 \times e^{-3}$, $5.0 \times e^{-4}$, $1.0 \times e^{-4}$, $5.0 \times e^{-5}$, $1.0 \times e^{-5}$, $5.0 \times e^{-6}$, $1.0 \times e^{-6}$\}. For a mini-batch size $B$, we  use 32 for GRU, and 64 or 1024 for NCDEs, depending on the dataset size. In the case of NCDEs, we choose Runge-Kutta (RK4) method as an ODE solver, and train them for 100 epochs, whereas train GRU for 300 epochs.

\subsection{Experimental Results}

In this section, we answer several research questions (RQs). Herein, we focus on the two standard metric scores of recommendation: Recall@20 and NDCG@20. For the evaluations of the time-series forecasting, we refer to RQ3.

\subsubsection {RQ1} \textbf{How accurate is the proposed method in recommending for the future?}
For answering RQ1, we first compare our method with the original scores of the base algorithms --- i.e., the base algorithms are trained with $\{(\mathcal{D}_i, t_i)\}_{i=1}^{M}$ and recommend for $(t_M, t_{M+1}]$. Table~\ref{tab:main_score} shows that our method outperforms the original base algorithms by large margins. Among various time-series forecasting methods, GRU-NCDE shows the best results in many cases, and GRU-GRU is the most effective in few cases.

For Gowalla dataset, LT-OCF shows the best scores and GRU-GRU significantly improves the scores by 31.57\% for Recall@20 and 37.86\% for NDCG@20. For other datasets, LightGCN is better than LT-OCF, and our TimeKit further improves LightGCN. 

The biggest enhancements are obtained in Netflix for NGCF with GRU-NCDE. It improves NGCF by 68.46\% and 68.75\% for Recall@20 and NDCG@20 respectively, achieving the best scores among all methods for Netflix. In general, GRU-NCDE is the best upgrade method except for Gowalla. 

For other cases, the improvement ratios are between 0.66\% to 47.69\%. Just adopting our upgrade kit, base algorithms' scores are considerably increased. Especially in the case of BPRMF with TimeKit, it even exceeds the original scores of both LightGCN and LT-OCF in Amazon-book --- Recall@20 of 0.0575 and NDCG@20 of 0.0303 by BPRMF with TimeKit vs. Recall@20 of 0.0573 and NDCG@20 of 0.0299 by LT-OCF. It also shows comparable scores in other datasets as well. These results indicate that TimeKit greatly improves BPRMF, the space/time efficient model, to better capture complicated relationships between users and items than CF models.

\begin{table*}[t]
\setlength{\tabcolsep}{4pt}
\centering
\scriptsize
\caption{Sensitivity w.r.t. the input time duration. We reduce the input information in comparison with the experiments in Table~\ref{tab:main_score}. Since the testing sample numbers also decrease, the direct comparison with Table~\ref{tab:main_score} is not appropriate.}
\label{tab:sensitivity_score}
\begin{tabular}{c|c|cc|cc|cc|cc|cc}
\toprule
\multirow{3}{*}{Dataset}   & \multirow{3}{*}{Base Alg. $\mathcal{A}$} & \multicolumn{2}{c|}{Original} & \multicolumn{2}{c|}{GRU-GRU}      & \multicolumn{2}{c|}{NCDE-NCDE} & \multicolumn{2}{c|}{GRU-NCDE}      & \multicolumn{2}{c}{Improvement (\%)} \\ \cmidrule{3-12} 
               &            & Recall@20       & NDCG@20      & Recall@20       & NDCG@20         & Recall@20      & NDCG@20     & Recall@20       & NDCG@20         & Recall@20         & NDCG@20    \\ \midrule
                                                                        
\multirow{4}{*}{Amazon-book}                                              
                            & BPRMF                   & 0.0542          & 0.0282       & 0.0752          & 0.0428          & 0.0769         & 0.0444      & \textbf{0.0770}  & \textbf{0.0448} & \underline{42.17}  & \underline{58.75}  \\ \cline{2-12} 
                            & NGCF                    & 0.0440          & 0.0225       & 0.0458          & 0.0239          & 0.0573         & 0.0318      & \textbf{0.0591}  & \textbf{0.0327} & 34.44              & 45.04              \\ \cline{2-12} 
                            & LightGCN   & \cellcolor[HTML]{BCBCBC}0.0661          & \cellcolor[HTML]{BCBCBC}0.0351       & 0.0701           & 0.0377           & 0.0790          & 0.0451             & \textbf{0.0814} & \textbf{0.0470} & 23.22    & 33.90  \\ \cline{2-12} 
                            & LT-OCF                  & 0.0654          & 0.0345       & 0.0671          & 0.0349          & 0.0803         & 0.0455      & \textbf{0.0837}  & \textbf{0.0478} & 28.07              & 38.54              \\ \hhline{============}
\multirow{4}{*}{Goodreads} 
                            & BPRMF                   & 0.0342          & 0.0200       & 0.0405          & 0.0245          & 0.0397         & 0.0238      & \textbf{0.0406}  & \textbf{0.0245} & \underline{18.71}  & \underline{22.68}  \\ \cline{2-12} 
                            & NGCF                    & 0.0295          & 0.0171       & 0.0300	         & 0.0177	       & 0.0319	& 0.0199	& \textbf{0.0326}	& \textbf{0.0205}          & 10.78	& 19.94            \\ \cline{2-12} 
                            & LightGCN                & \cellcolor[HTML]{BCBCBC}0.0445                   & 0.0259          & 0.0470         & 0.0276      & 0.0489           & 0.0290          & \textbf{0.0506}    & \textbf{0.0300} & 13.65    & 15.79   \\ \cline{2-12} 
                            & LT-OCF                  & 0.0444          & \cellcolor[HTML]{BCBCBC}0.0263       & 0.0445    & 0.0265         & 0.0513      & 0.0315           & \textbf{0.0523} & \textbf{0.0319}    & 17.91             & 21.70  \\ \bottomrule

\end{tabular}
\end{table*}

\subsubsection {RQ2} \textbf{How accurate is the proposed method when we reduce the input information?}
In Table~\ref{tab:main_score}, we used the full period of data. For answering RQ2, we remove the early 25\% of periods
from Amazon-Book and Goodreads, using only newer interactions --- other datasets have a relatively short period of data and are excluded from this study. While filtering out the early 25\%, we reconstruct the training/validating/testing sets and therefore, the direct comparison between Tables~\ref{tab:main_score} and~\ref{tab:sensitivity_score} is not possible. 

As shown in Table~\ref{tab:sensitivity_score}, our upgrade kit is still effective. Overall, GRU-NCDE is the best upgrade kit, leading to the biggest scores in all cases and achieving the improvement ratio of up to 58.75\%. This result also implies that Timekit outperforms baseline algorithms that are trained with relatively recent interactions.

\begin{table}[t]
\setlength{\tabcolsep}{3pt}
\centering
\scriptsize
\caption{MSE of GRU-NCDE. We also include the absolute value of the maximum (M) minus the minimum (m), and the std. dev. of the elements in the user and item embedding vectors.}
\label{tab:mse_loss}
\begin{tabular}{c|c|ccc|cc|ccc}
\toprule
\multirow{3}{*}{Dataset} & \multirow{3}{*}{Model} &
\multicolumn{3}{c|}{GRU MSE loss}&\multicolumn{2}{c|}{Statistics}&\multicolumn{3}{c}{NCDE MSE loss}\\\cmidrule{3-10}
& & Train & Valid & Test & $|$M-m$|$ & Std &Train & Valid & Test  \\ 
\midrule 
\multirow{4}{*}{\rotatebox{90}{Gowalla}}    
                             & BPRMF        & 0.033& 0.032& 0.029       & 4.39   &0.33      &  0.012 & 0.011    &0.009      \\ \cline{2-10} 
                             & NGCF         & 0.008& 0.007& 0.044       & 3.96   &0.28      &  0.006 & 0.005    &0.021       \\ \cline{2-10} 
                             & LightGCN     & 0.040& 0.068& 0.044       & 4.24   &0.22      &  0.003 & 0.006    &0.006      \\ \cline{2-10} 
                             & LT-OCF       & 0.003& 0.536& 0.006       & 4.33   &0.19      &  0.003 & 0.012    &0.012    \\ \hhline{==========}
\multirow{4}{*}{{\rotatebox{90}{\begin{tabular}[c]{@{}c@{}}Amazon-\\Book\end{tabular}}}}
                             & BPRMF        & 0.058& 0.058& 0.048       & 4.39   &0.42      &  0.042 & 0.042    &0.030     \\ \cline{2-10}
                             & NGCF         & 0.047& 0.055& 0.092       & 6.98   &0.40      &  0.035 & 0.046    &0.066   \\ \cline{2-10} 
                             & LightGCN     & 0.031& 0.030& 0.022       & 4.43   &0.35      &  0.021 & 0.025    &0.017   \\ \cline{2-10}
                             & LT-OCF       & 0.054& 0.020& 0.022       & 3.53   &0.35      &  0.022 & 0.028    &0.018  \\ \hhline{==========}
\multirow{4}{*}{{\rotatebox{90}{\begin{tabular}[c]{@{}c@{}}Good-\\reads\end{tabular}}}}  
                             & BPRMF        & 0.091& 0.086& 0.080       & 3.82   &0.38      &  0.034 & 0.033    &0.028    \\ \cline{2-10}
                             & NGCF         & 0.145& 0.159& 0.236       & 8.15   &0.36      &  0.048 & 0.055    &0.091   \\ \cline{2-10}
                             & LightGCN     & 0.040& 0.040& 0.035       & 5.00   &0.31      &  0.019 & 0.021    &0.017 \\ \cline{2-10}
                             & LT-OCF       & 0.052& 0.052& 0.043       & 2.70   &0.31      &  0.021 & 0.023    &0.018  \\ \hhline{==========}
\multirow{4}{*}{\rotatebox{90}{Netflix}}   
                             & BPRMF        & 0.085& 0.088& 0.098       & 4.49  &0.41       &  0.023 & 0.037    &0.044  \\ \cline{2-10}
                             & NGCF         & 0.001& 0.010& 0.048       & 10.55 &0.75       &  0.002 & 0.017    &0.058\\ \cline{2-10}
                             & LightGCN     & 0.008& 0.012& 0.022       & 6.88  &0.27       &  0.003 & 0.006    &0.017  \\ \cline{2-10}
                             & LT-OCF       & 0.002& 0.005& 0.013       & 5.38  &0.26       &  0.002 & 0.005    &0.013   \\ 
\bottomrule
\end{tabular}
\end{table}

\begin{table*}[t]
\caption{Empirical complexity analysis}
\setlength{\tabcolsep}{3pt}
\centering
\scriptsize
\label{tab:mem_time}
\begin{tabular}{c|cccccc|cccccc}
\toprule
\multirow{4}{*}{Dataset} & \multicolumn{6}{c|}{Memory (MB)}                                                            & \multicolumn{6}{c}{Training Time (seconds per epoch)}                                                     \\ \cmidrule{2-13} 
                        & \multicolumn{3}{c|}{User embedding}              & \multicolumn{3}{c|}{Item embedding} & \multicolumn{3}{c|}{User embedding}              & \multicolumn{3}{c}{Item embedding} \\ \cmidrule{2-13} 
                        & GRU & NCDE(RK4) & \multicolumn{1}{c|}{NCDE(DOPRI)} & GRU    & NCDE(RK4)    &  NCDE(DOPRI)   & GRU & NCDE(RK4) & \multicolumn{1}{c|}{NCDE(DOPRI)} & GRU    & NCDE(RK4)   & NCDE(DOPRI)   \\ \midrule
Gowalla                 &  193   &	217	&  \multicolumn{1}{c|}{302}    & 4,085 & 1,570 &	1,218 &	2.06 &	2.29 & \multicolumn{1}{c|}{42.74}            &    2.92 & 	11.96 &	381.76      \\ 
Amazon-book             &  805   &	376 &  \multicolumn{1}{c|}{326}    & 2,148 & 837   &	692   &	2.18 &	3.59 & \multicolumn{1}{c|}{129.23}           &    3.38 &	7.92  &	281.49      \\ 
Goodreads               &  805   &	375 &  \multicolumn{1}{c|}{326}    & 4,707 & 1,719 &	1,388 &	1.25 &	3.57 & \multicolumn{1}{c|}{159.69}           &    3.21 &	15.77 &	661.64      \\ 
Netflix                 &  1,628 &  712 &  \multicolumn{1}{c|}{560}    & 1,266 & 577   &	458   &	2.22 &	4.45 & \multicolumn{1}{c|}{203.38}           &    2.18 &	3.89  &	248.25      \\ \bottomrule
\end{tabular}
\end{table*}

\subsubsection {RQ3} \textbf{How well does the proposed method extrapolate (or forecast) embedding vectors?}
Table~\ref{tab:mse_loss} summarizes the MSE values of our GRU-NCDE forecasting model, i.e., GRU-based models for user embeddings and NCDE-based models for item embeddings, which generally shows the best performance. In many cases, our testing MSEs are similar to those of training and validating, which means stable forecasting. In addition, the MSE values are one or two orders of magnitude smaller than the std. dev. of item embedding elements. From all these facts, we can know that our forecasting is reliable.

\begin{figure}
    \centering
    \includegraphics[width=0.9\columnwidth]{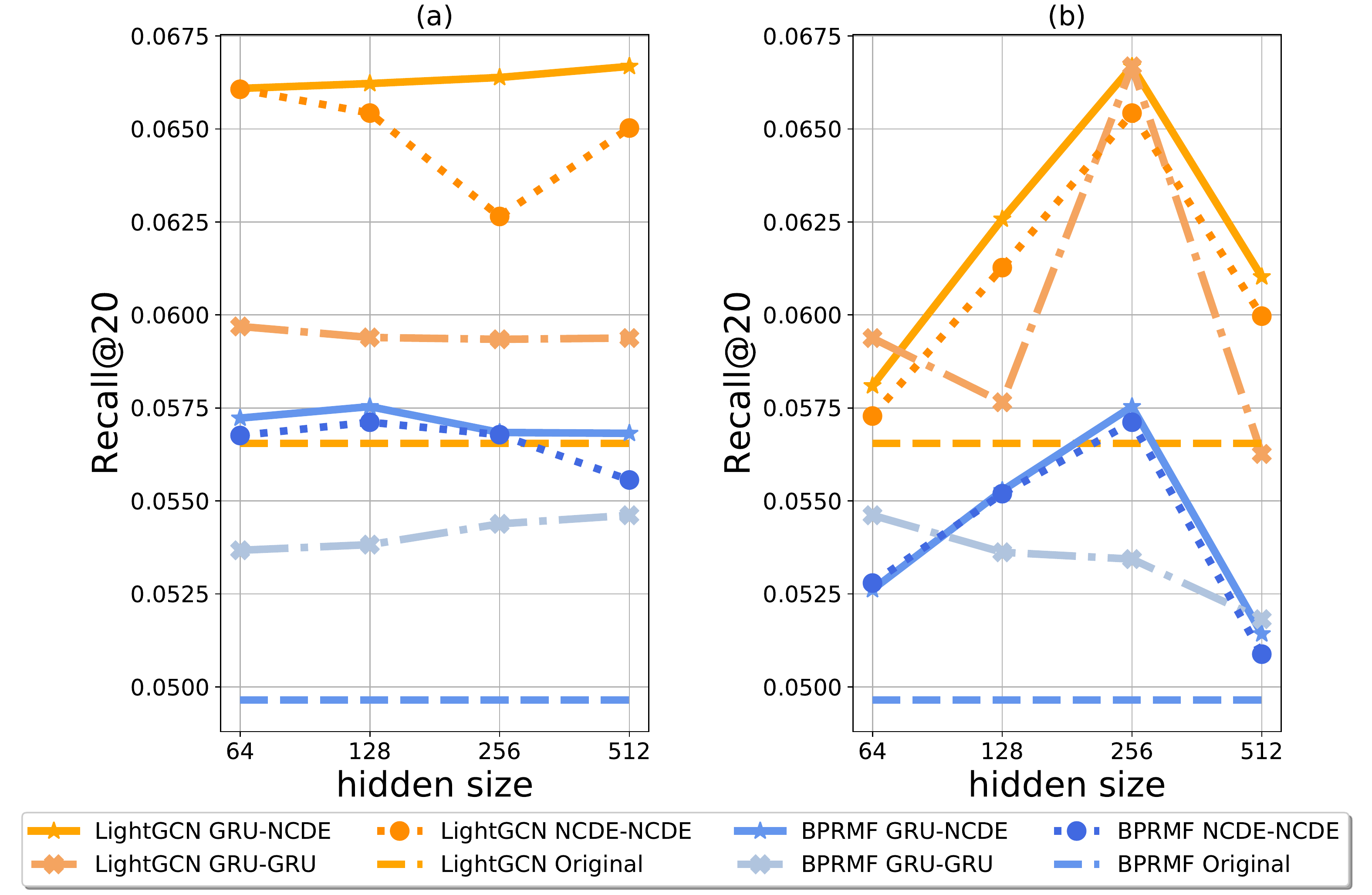}
    \caption{{Sensitivity to the hidden size on Amazon-book. (a) We show the Recall@20 scores by changing the hidden size of the embedding forecaster for users and fixing it for items. (b) We fix the hidden size for users and change for items. As noted, the item embedding vectors are more sensitive to the hyperparameter.}}
    \label{fig:hid_sens}
\end{figure}

\begin{figure}[t]
    \centering
    \subfigure[Only our method, which is trained with the latent dynamics of the preference distribution changes over time, correctly forecasts that User A will watch those family animations and User B will watch those romantic comedy films at December 2005 whereas base algorithms, trained with the ``Total'' distributions, fail to do so.]{\includegraphics[width=\columnwidth]{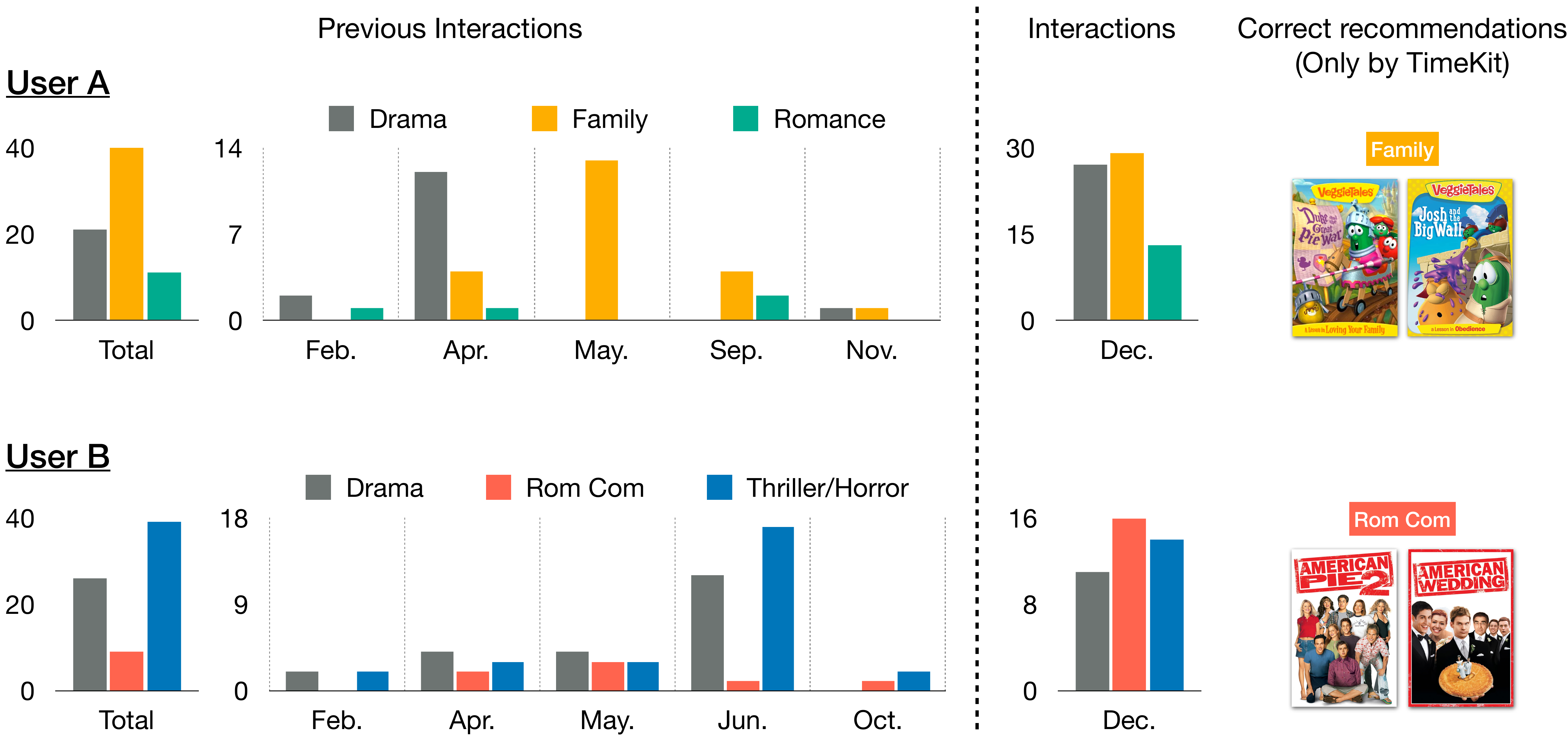}}
    \subfigure[User A's embedding vectors]{\includegraphics[width=0.49\columnwidth]{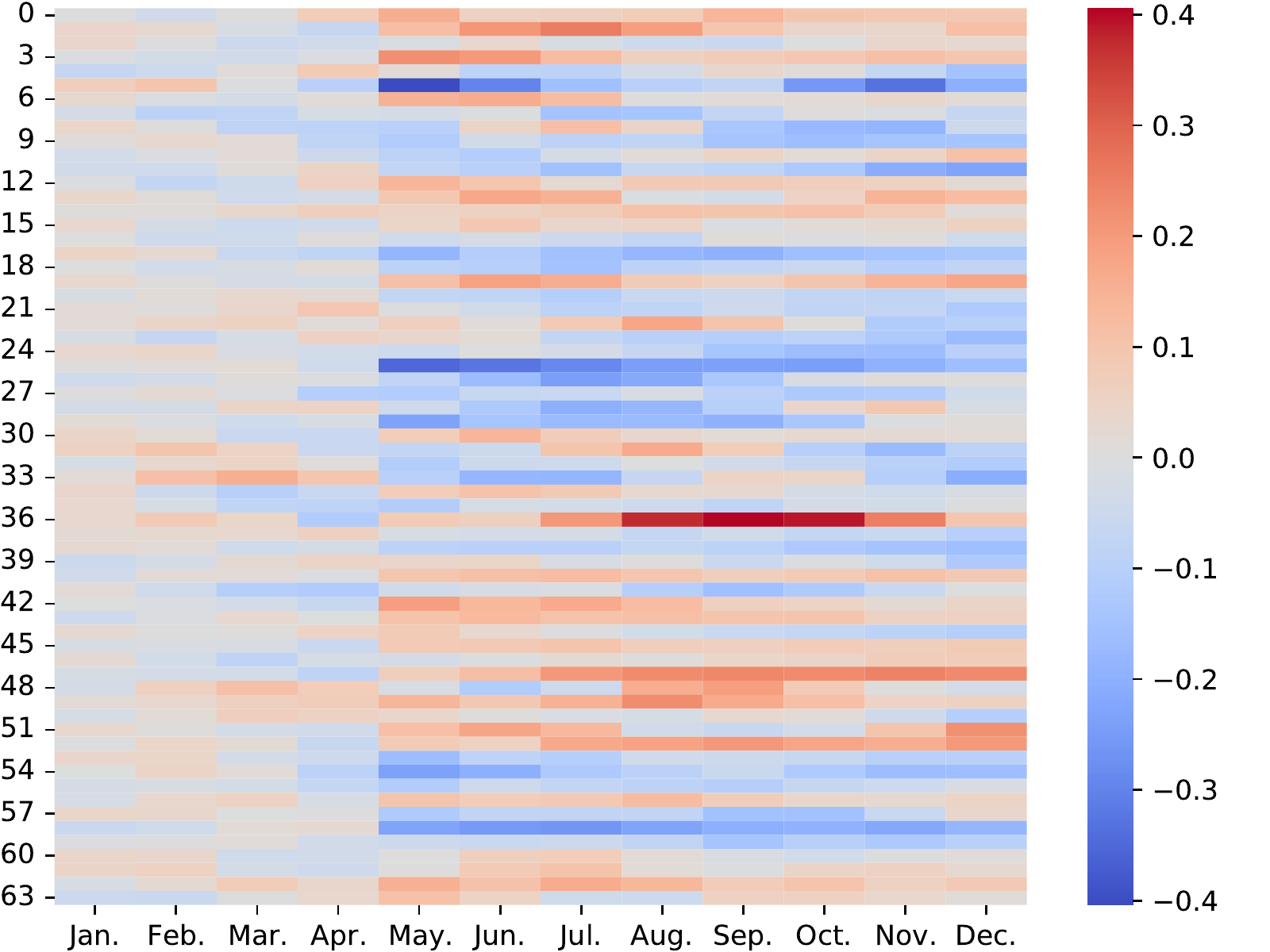}}
    \subfigure[User B's embedding vectors]{\includegraphics[width=0.49\columnwidth]{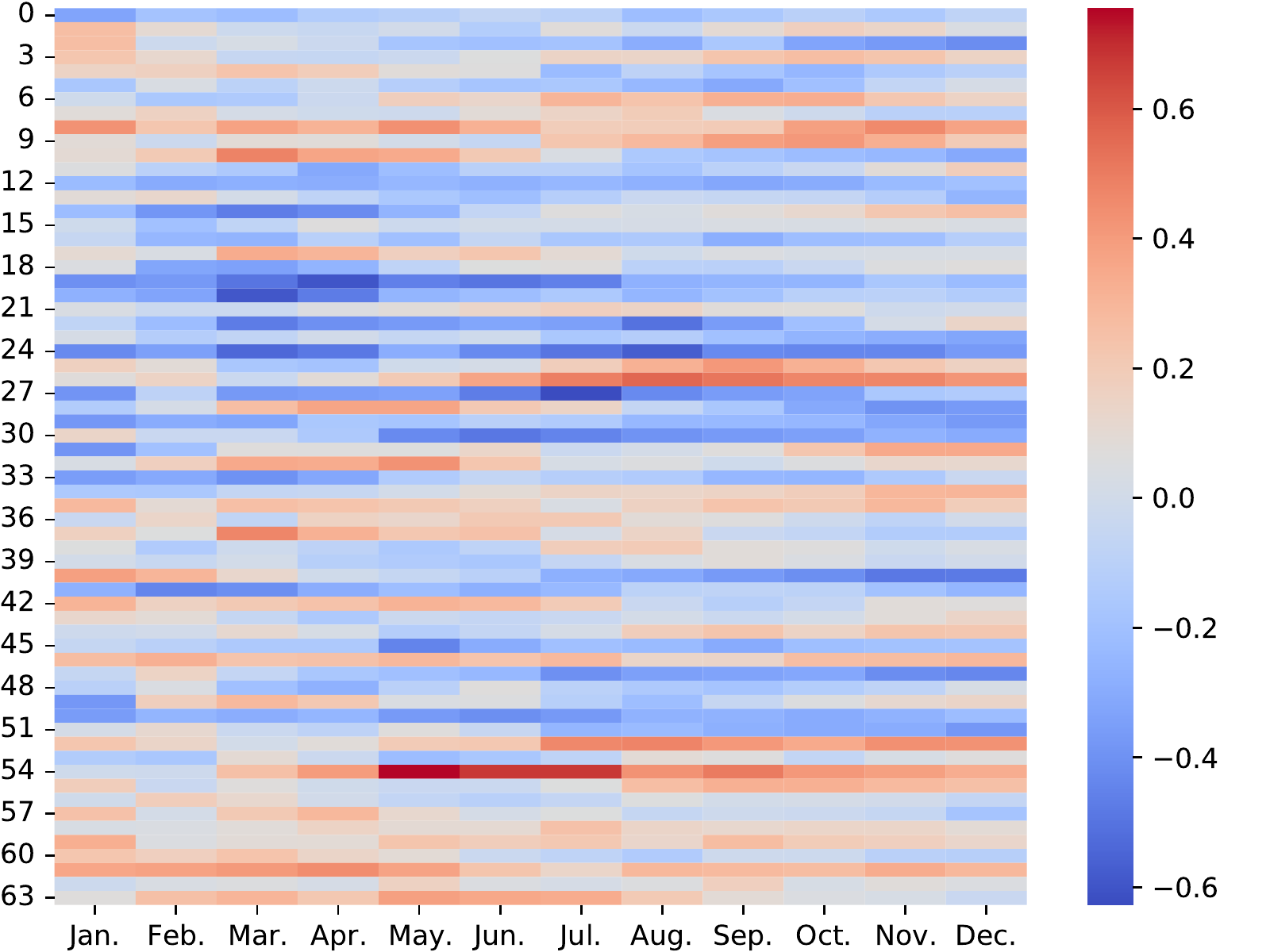}}
    
    \caption{Two users' recommendation examples and embedding vectors change over time in Netflix using LightGCN. We note that the preference distribution of User A in (a) abruptly changes at May and so are the embeddings in (b).}\label{fig:film_dynamics}
\end{figure}

\subsubsection {Sensitivity Analysis} \textbf{How accurate is the proposed method when we change the hidden size?}
Fig.~\ref{fig:hid_sens} shows the sensitivity analysis result. For various hidden size settings of our embedding forecasters, our method still shows better performances than base algorithms in all but one case. Even though the hidden size changes, TimeKit-applied BPRMF outperforms LightGCN's original scores in some cases, which proves the efficacy of TimeKit. 




{
\subsubsection {Empirical Training Complexity Analysis} \textbf{What is the empirical training complexity of our upgrade kit?}
We analyze the empirical training overhead incurred by our method since we need to additionally train time-series forecasting models. As shown in Table~\ref{tab:mem_time}, the empirical space and time complexities of our proposed upgrade kit are not large. For training our GRU or NCDE-based forecasting models for user embedding vectors, it requires just hundreds of megabytes for GPU memory and takes a few seconds for each epoch (except the NCDE-based model with DOPRI). However, all of our experimental results were obtained with RK4 as the performance enhancement by DOPRI is neglectable. 

\subsubsection {Case Study \& Visualization}
We introduce some key observations from our experiments, which intuitively explains how our method works. In Fig.~\ref{fig:film_dynamics} (a), our method exclusively and correctly forecasts some family animations for User A and some romantic comedy films for User B, but two users show different patterns. User A suddenly watched lots of family movies in May, and this preference was reflected in the embedding (cf. Fig.~\ref{fig:film_dynamics} (b)). User B, on the contrary, does not show a strong preference on romantic comedy films during his/her training period but a consistent preference to some degree. However, the base algorithm cannot capture those differences because what they learn is not the dynamics how the distributions change but the ``Total'' distribution in Fig.~\ref{fig:film_dynamics} (a). Because of that difference, only our method correctly forecasts the phenomena that they show strong preferences on family and romantic comedy movies in December. Rating scores of the correct items are also the highest in December, as indicated in Fig~\ref{fig:score_dynamics}. 

\begin{figure}[t]
\centering 
\includegraphics[width=0.8\columnwidth]{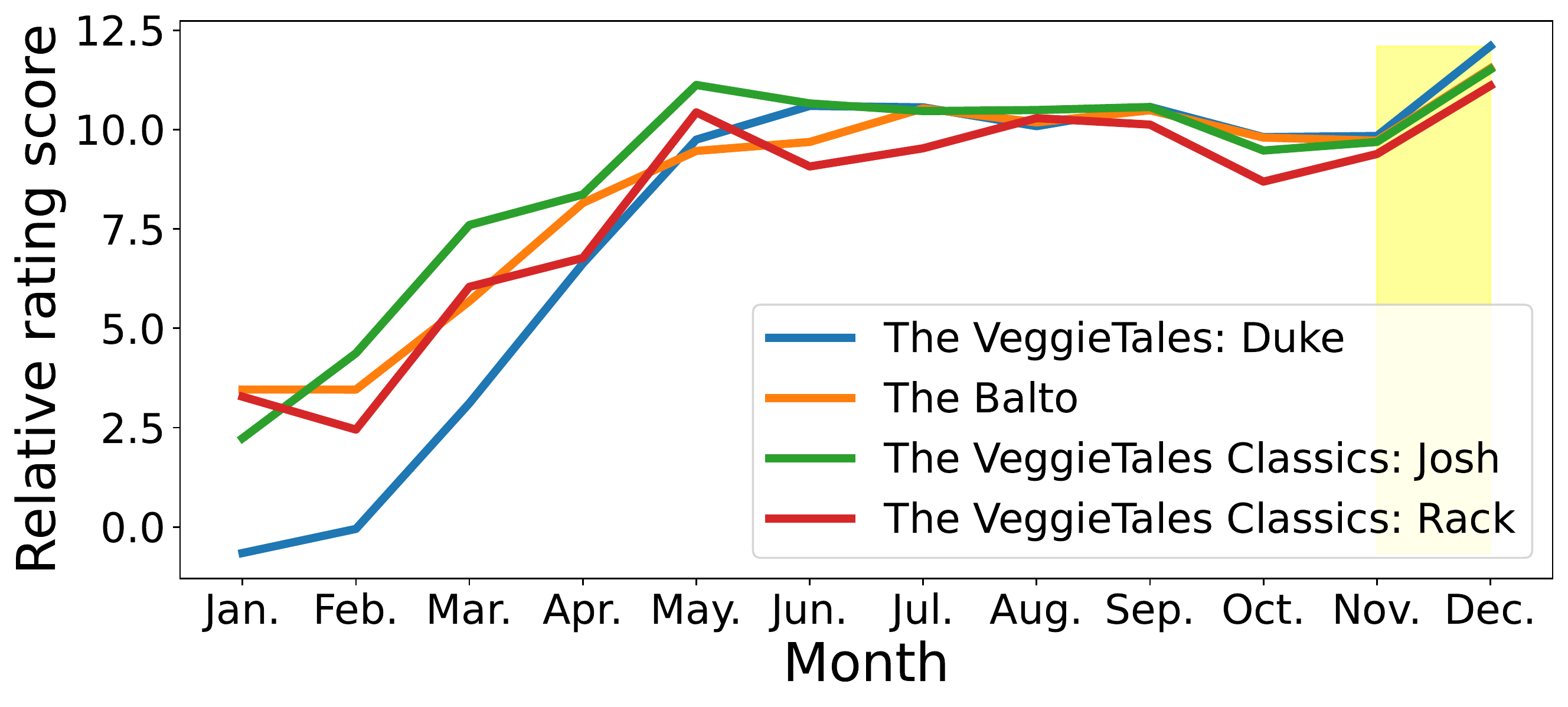}
\caption{The dynamics of films' relative rating scores for User A. The relative rating score is the ratio of a single item's rating score (user-item dot-product value with TimeKit) to the average of all rating scores, which shows how much each film is preferred in comparison with the average.}
\label{fig:score_dynamics}
\end{figure}

\section{Conclusions and Limitations}
We presented a novel upgrade kit, called TimeKit. Our goal is to forecast the future user/item embedding vectors, with which we will perform the collaborative filtering task. In other words, we reduce the collaborative filtering to the time-series forecasting task, owing to a recent technological breakthrough in dealing with complicated time-series data. In general, user/item embedding vectors are known to contain the hidden collaborative information for the recommendation. Therefore, we need to uncover the hidden dynamics to accurately forecast the future embedding vectors. NCDEs are a powerful concept in unveiling a hidden dynamics from time-series and we resort to this technique. We conduct experiments with four benchmark datasets and four base collaborative filtering algorithms. All those base algorithms are significantly improved when being upgraded with TimeKit.


One limitation is that we need to collect time-series data during enough periods. In the real world, however, the model is incrementally updated as new interactions are added, and the time-series data occurs naturally. Moreover, user/item embeddings utilize less memory, making them less of a burden for capacity. Overall, our proposed method is lightweight and does not necessitate much training or a large memory. 

}

\section*{Acknowledgment}
Noseong Park is the corresponding author. This work was supported by an IITP grant funded by the Korean government (MSIT) (No.2020-0-01361, Artificial Intelligence Graduate School Program (Yonsei University)) and an ETRI grant funded by the Korean government (22ZS1100, Core Technology Research for Self-Improving Integrated Artificial Intelligence System).

\bibliographystyle{IEEEtran}
\bibliography{sample-base}

\end{document}